\documentclass[aps,prb,twocolumn,showpacs,floatfix]{revtex4-1}
\usepackage{amsmath,amsfonts,amssymb}
\usepackage{multirow}
\usepackage{xspace,units}
\usepackage[english]{babel}
\usepackage{multirow}
\usepackage{natbib}

\usepackage{graphicx,color,fancyhdr,xspace}
\definecolor{darkgreen}{rgb}{0,0.5,0}
\definecolor{darkblue}{rgb}{0,0,0.2}
\definecolor{purple}{rgb}{0.35,0,0.35}
\definecolor{orange}{rgb}{1,0.5,0}

\RequirePackage[
   hyperindex,colorlinks,bookmarksnumbered,
   plainpages=true,pdfstartview=FitH]{hyperref}
\hypersetup{linkcolor=blue,urlcolor=darkblue,citecolor=darkgreen}
\usepackage[pdfstartview=FitH]{hyperref}

  \newcommand{\Sec}[1]{Sec.~\ref{#1}}
  
  \newcommand{\App}[1]{App.~\ref{#1}}

  \newcommand{\Eq}[1]{Eq.\,(\ref{#1})}
  \newcommand{\Eqs}[1]{Eqs.\,(\ref{#1})}
  \newcommand{\Eqr}[2]{Eqs.\,(\ref{#1}-\ref{#2})}

  \newcommand{\FIG}[1]{Figure~\ref{#1}}

  \newcommand{\Fig}[1]{Fig.~\ref{#1}}
  
  \newcommand{\Figsp}[2]{Figs.~\ref{#1}(#2)}

  \newcommand{\Tbl}[1]{Tbl.~\ref{#1}}

  \newcommand{\ie}{i.e.\xspace}
  \newcommand{\eg}{e.g.\xspace}
  \newcommand{\vs}{vs.\xspace}
  \newcommand{\wrt}{w.r.t.\xspace}
  \newcommand{\cf}{cf.\xspace}

  \newcommand{\Hc}{\ensuremath{\mathrm{H.c.}}}

  \newcommand{\D}{\ensuremath{\mathrm{D}}}

  \newcommand{\Etrunc}{\ensuremath{E_\mathrm{tr}}\xspace}

  \newcommand{\imp}{\ensuremath{\mathrm{d}}\xspace}
  \newcommand{\impb}{\ensuremath{\mathrm{(d)}}\xspace}
  \newcommand{\Sz}{\ensuremath{\hat{S}_z}\xspace}
  \newcommand{\Szimp}{\ensuremath{\hat{S}_z^{\mathrm{d}}}\xspace}
  \newcommand{\Sztotv}{\ensuremath{S_z^{\mathrm{tot}}}\xspace}
  \newcommand{\Sztot}{\ensuremath{\hat{S}_z^{\mathrm{tot}}}\xspace}
  \newcommand{\Szbath}{\ensuremath{\hat{S}_z^{\mathrm{bath}}}\xspace}

  \newcommand{\Himp}{\ensuremath{\hat{H}_{\mathrm{imp}}}\xspace}
  \newcommand{\Hcpl}{\ensuremath{\hat{H}_{\mathrm{cpl}}}\xspace}
  \newcommand{\Hbath}{\ensuremath{\hat{H}_{\mathrm{bath}}}\xspace}

  \newcommand{\Hipc}{\ensuremath{\hat{H}_{\mathrm{ipc}}}\xspace}
  \newcommand{\Tipc}{\ensuremath{T_{\mathrm{ipc}}}\xspace}
  \newcommand{\Hbpc}{\ensuremath{\hat{H}_{\mathrm{bpc}}}\xspace}
  \newcommand{\Tbpc}{\ensuremath{T_{\mathrm{bpc}}}\xspace}
  \newcommand{\Htot}{\ensuremath{\hat{H}_{\mathrm{tot}}}\xspace}
  \newcommand{\Htotb}{\ensuremath{\hat{H}_{\mathrm{(tot)}}}\xspace}
  \newcommand{\Ttot}{\ensuremath{T_{\mathrm{(tot)}}}\xspace}

  \newcommand{\TK}{\ensuremath{T_\mathrm{K}}\xspace}
  \newcommand{\TKd}{\ensuremath{T_\mathrm{K}^\mathrm{d}}\xspace}
  \newcommand{\TKfs}{\ensuremath{T_\mathrm{K}^\mathrm{FS}}\xspace}
  \newcommand{\TKsc}{\ensuremath{T_\mathrm{K}^\mathrm{sc}}\xspace}
  \newcommand{\TKscB}{\ensuremath{T_\mathrm{K}^\mathrm{sc,B}}\xspace}
  \newcommand{\TKscx}{\ensuremath{T_\mathrm{K}^{\mathrm{sc},x}}\xspace}
  \newcommand{\TKphi}{\ensuremath{T_\mathrm{K}^{\varphi}}\xspace}
  \newcommand{\TKinf}{\ensuremath{T_\mathrm{K}^{\infty}}\xspace}

  \newcommand{\THg}{\ensuremath{T_{\nicefrac{1}{2}}}\xspace}
  \newcommand{\THd}{\ensuremath{T_{\nicefrac{1}{2}}^\mathrm{d}}\xspace}
  \newcommand{\THsc}{\ensuremath{T_{\nicefrac{1}{2}}^\mathrm{sc}}\xspace}

  \newcommand{\BHg}{\ensuremath{B_{\nicefrac{1}{2}}}\xspace}
  \newcommand{\BHd}{\ensuremath{B_{\nicefrac{1}{2}}^\mathrm{d}}\xspace}
  \newcommand{\BHsc}{\ensuremath{B_{\nicefrac{1}{2}}^\mathrm{sc}}\xspace}

  \newcommand{\chic}{\ensuremath{\chi^\mathrm{c}}\xspace} 
  \newcommand{\chid}{\ensuremath{\chi^\mathrm{d}}\xspace}
  \newcommand{\chisc}{\ensuremath{\chi^\mathrm{sc}}\xspace}
  \newcommand{\chiscB}{\ensuremath{\chi^\mathrm{sc,B}}\xspace}
  \newcommand{\chitot}{\ensuremath{\chi^\mathrm{tot}}\xspace}
  \newcommand{\chifs}{\ensuremath{\chi^\mathrm{FS}}\xspace}
  \newcommand{\chiinf}{\ensuremath{\chi^{\infty}}\xspace}

  \newcommand{\chiXY}{\ensuremath{\chi_{_\mathrm{XY}}}^\mathrm{R}\xspace}

  \newcommand{\nlocds}{\ensuremath{\langle \hat{n}_{\mathrm{loc},\sigma}(\varepsilon_{d})\rangle\xspace}}

  \def\myempty{myemptytoken}
  \newcommand{\SPTn}[3][\myempty]{
     \ifx\myempty#1\ensuremath{\langle #2 \Vert #3 \rangle_T}\xspace
     \else\ensuremath{\langle #2 \Vert #3 \rangle_T^{(#1)}}\xspace
     \fi
  }
  \newcommand{\SPTh}[3][\myempty]{
     \ifx\myempty#1\ensuremath{\langle\hat{#2}\Vert\hat{#3}\rangle_T}\xspace
     \else\ensuremath{\langle\hat{#2}\Vert\hat{#3}\rangle_T^{(#1)}}\xspace
     \fi
  }

\begin{document}

\title{Local susceptibility and Kondo scaling
in the presence of finite bandwidth}

\author{Markus \surname{Hanl}}
\affiliation{Physics Department, Arnold Sommerfeld Center for
Theoretical Physics, and Center for NanoScience,
Ludwig-Maximilians-Universit\"at, 80333 Munich, Germany }

\author{Andreas \surname{Weichselbaum}}
\affiliation{Physics Department, Arnold Sommerfeld Center for
Theoretical Physics, and Center for NanoScience,
Ludwig-Maximilians-Universit\"at, 80333 Munich, Germany }

\begin{abstract}
  The Kondo scale \TK for impurity systems is expected to
  guarantee universal scaling of physical quantities. However, in
  practice, not every definition of \TK necessarily supports this
  notion away from the strict scaling limit. Specifically, this
  paper addresses the role of finite bandwidth $D$ in the
  strongly-correlated Kondo regime. For this, various theoretical
  definitions of \TK are analyzed based on the inverse magnetic
  impurity susceptibility at zero temperature. While conventional
  definitions in that respect quickly fail to ensure universal
  Kondo scaling for a large range of $D$, this paper proposes an altered
  definition of \TKsc that allows universal scaling of dynamical
  or thermal quantities for a given fixed Hamiltonian. If the
  scaling is performed with respect to an external parameter which
  directly enters the Hamiltonian, such as magnetic field, the
  corresponding \TKscB for universal scaling differs, yet becomes
  equivalent to \TKsc in the scaling limit. The only requirement
  for universal scaling in the full Kondo parameter regime with a
  residual error of less than 1\% is a well-defined isolated Kondo
  feature with $\TK\lesssim 0.01\,D$, irrespective of specific other
  impurity parameter settings. By varying $D$ over a wide
  range relative to the bare energies of the impurity, for
  example, this allows a smooth transition from the Anderson to
  the Kondo model.
%
\end{abstract}

\date{\today}



\maketitle

\section{Introduction \label{sec:intro}}

The Kondo scale represents a dynamically generated low-energy scale
which arises when an unpaired spin, to be referred to as the
impurity, is screened by a metallic host. Prototypical examples
include actual dilute magnetic impurities in metals, \cite{Kondo64,
Bauerle05,Mallet06,Costi09} but also highly controllable quantum dot
settings which are characterized through transport measurements.
\cite{GoldhaberGordon98,Kretinin11} The precise definition of the
Kondo scale, however, is usually subject to conventions. 
Nevertheless, whatever the definition of the Kondo scale \TK, clean
isolated Kondo features are expected to be universal: that is after
proper scaling \wrt \TK, the resulting data is expected to fully
collapse onto a single universal curve. Therefore whatever the
specific definition of the Kondo scale, \eg up to an irrelevant
definition-dependent prefactor of order one, this represents an
important stringent requirement: \TK must allow for accurate scaling
of Kondo related features.
A prototypical application that requires such scaling, for example,
is the analysis of the prefactors in Fermi-liquid scaling of
interacting impurity models, \cite{Nozieres74, Grobis08, 
Pletyukhov12, Merker13} which strongly depends on the precise
definition of \TK. As a matter of fact, the present work emerged and
thus was motivated from preliminary work in exactly this direction
for multi-band models, \cite{Costi09,Hanl13} with the results on the
related Fermi liquid coefficients to be published elsewhere.

With \TK typically described by an exponential expression,
\cite{Hewson93} the terms in the exponent usually do not depend on
the full bandwidth $D$ of a given model. The prefactor in the
definition of \TK, however, may depend on $D$ with the consequence
that certain definitions of \TK can spoil universal Kondo scaling
\emph{even if} $\TK\ll D$.
Consider, for example, the standard single impurity Anderson model 
(SIAM, see model Hamiltonian further below) with the impurity onsite 
interaction $U$. For $U\ll D$ the full bandwidth $D$ becomes 
irrelevant for the impurity related physics. This turns out to be the 
safe regime for impurity related quantities. For the case $U\gtrsim 
D$, however, the bandwidth $D$ becomes relevant for Kondo related 
quantities. Importantly, this regime is (i) experimentally relevant, 
in that the experiment is never truly in the Kondo scaling limit. 
Moreover, through Schrieffer-Wolff transformation in the limit 
$U\to\infty$ of the particle-hole symmetric SIAM, (ii) this leads to 
the Kondo model, a widely used model itself. With its Kondo 
temperature given by $\TK \simeq D \sqrt{2 \nu J} e^{-1/(2\nu J)}$, 
\cite{Kondo64,Hewson93,Costi00} with $J$ the Kondo coupling and
$\nu$ the density of states at the Fermi edge, this model is
\emph{intrinsically and strongly} affected by finite bandwidth.
Therefore, in particular, the present discussion is of clear
relevance also for the Kondo model.

Proper Kondo scaling is already built-in by construction in the 
experiment-like approach of using (full-\mbox{width-)}\,half-maximum 
type measures of \TK, \cite{GoldhaberGordon98,Kretinin11} which 
strictly focuses on the low-energy features of the measured 
quantities, typically assuming $\TK\ll D$. However, this requires to 
measure or calculate an entire curve while possibly subtracting a 
broader background still.\cite{Costi09} In contrast, for the 
theoretical analysis it appears more desirable to have a single 
measurable quantity, instead, which uniquely defines \TK up to a 
convention-dependent constant prefactor of order one. To be specific, 
this requires a definition of \TK at zero temperature in the absence 
of magnetic field in a static context, \ie $T=B=\omega=0$ (using 
$k_B=g\mu_B=\hbar=1$ throughout, for convenience). This \TK is 
measured through a weak perturbation of the system, and hence can be 
computed within linear response. Considering that the Kondo state is 
sensitive to an external magnetic field, the quantity of interest 
discussed in this paper is the magnetic susceptibility of the 
impurity. The following discussion, however, can be generalized to 
other local susceptibilities. 

A standard definition for the Kondo temperature for the one-channel 
Kondo model is given by \cite{Hewson93,Bulla08}
\begin{equation}
   \TK \equiv \tfrac{1}{4\chi_0}
\text{,}  \label{def:TK:chi0}
\end{equation}
with $\chi _{0}\equiv \lim_{T\rightarrow 0}\chi \left( T\right) $ the 
static magnetic susceptibility of the impurity in the limit of zero 
temperature. The constant prefactor of $1/4$ is part of the 
definition which may be chosen differently, for example, for 
multi-channel models.\cite{Hewson93} The immanent question, however, 
that arises with \Eq{def:TK:chi0} is, how does one precisely define 
the impurity contribution $\chi _{0}$ to the magnetic susceptibility? 
The predominant conventions to be found in the literature are, 
\cite{Hewson93,Bulla08,Dirks13,Hoeck13} 
\begin{subequations}
\label{def:chi0:1}
\begin{align}
   \chi^\impb(T) & \equiv
   \langle \Szimp \Vert \Szimp \rangle _{T}
\label{def:chi0:Szimp} \\
   \chitot(T) & \equiv
   \langle \Sztot \Vert \Sztot \rangle _{T}
 - \langle \Sztot \Vert \Sztot \rangle _{T}^{\left( 0\right) }
\text{.}  \label{def:chi0:Sztot}
\end{align}
\end{subequations}
where $\langle \hat{S}^{\alpha}\Vert \hat{S}^{\beta}\rangle \equiv 
\bigl. \tfrac{d}{dB} \langle \hat{S}^{\beta}\rangle \bigr|_{B=0}$ 
describes the static linear spin susceptibility of $\langle 
\hat{S}^{\beta}\rangle$ in response to the perturbation 
$\hat{H}^{\prime} =-B\hat{S}^{\alpha}$ with $B$ an external magnetic 
field (the minus sign in $\hat{H}'$ ensures $\chi \ge 0$ if 
$\hat{S}^{\alpha} =\hat{S}^{\beta}$). Here $\hat{S}_z^\mathrm{d}$ 
($\hat{S}_z^\mathrm{tot}$) stands for the total spin of the impurity 
(the entire system), respectively. Since, in general, the	 spin of 
the impurity $\Szimp$ is not conserved and hence does not commute 
with the Hamiltonian, \Eq{def:chi0:Szimp} is equivalent to the 
evaluation of a dynamical correlation function. \cite{Bulla08} It is 
a somewhat abstract quantity since from an experimental point of view 
it is difficult to just apply a magnetic field at the impurity 
itself. The second definition of the impurity susceptibility in 
\Eq{def:chi0:Sztot}, on the other hand, is typically considered 
closer to an experimental realization, in that the impurity 
contribution to the total susceptibility is evaluated by taking the 
difference of the total susceptibility with [$\left\langle \cdot 
\right\rangle _{T}$]\ and without [$\left\langle \cdot \right\rangle 
_{T}^{\left( 0\right) }$] the impurity, where the latter acts as a 
reference system. \Eq{def:chi0:Sztot} includes the total spin 
$\Sztot$ of the system, which is assumed to be conserved and hence is 
simply proportional to the overall spin fluctuations, $\langle \Sztot 
\Vert \Sztot \rangle _{T}=\beta (\langle (\Sztot)^{2}\rangle -\langle 
\Sztot \rangle ^{2})$ where $\beta \equiv 1/T$. Hence, in principle, 
it is easier to evaluate. However, from a computational point of view 
it has the disadvantage that one essentially needs two calculations, 
one with and one without the impurity, followed by the subtraction of 
two extensive macroscopic and thus large values in order to obtain an 
intrinsic impurity-related finite quantity. While one may expect that 
both definitions in \Eqs{def:chi0:1} give comparable results, they 
are not strictly equivalent. In particular, neither definition in 
\Eqs{def:chi0:1} necessarily guarantees proper scaling of Kondo 
related features at finite bandwidth. 

Scaling onto a universal curve requires an appropriate and consistent 
set of parameters. For the Kondo physics analyzed in this paper, 
these are simply a particle-hole symmetric setting (or a similarly 
consistent asymmetric setting, \eg $U/\varepsilon_d=\mathrm{const}$ 
for the SIAM below), together with the bare requirement of a 
well-defined isolated low-energy feature with $\TK\lesssim 0.01\,D$, 
\eg the Kondo peak in the spectral function, which allows to observe 
Kondo physics to start with. 
Here universal scaling is understood in the usual way. Given a set of 
individual curves $y(x;\{p\})$, when plotted \vs $x$, these depend on 
a set $\{p\}$ of external model parameters. Here $x$ represents an 
energy, \eg $x\in \{\omega,T,B,\ldots\}$. Therefore universal scaling 
of $x$ by an appropriately chosen Kondo scale \TKscx, \ie $\tilde{x} 
\equiv x/\TKscx$, implies that the curves $y(\TKscx \tilde{x};\{p\}) 
/ y_0 =: \tilde{y}(\tilde{x})$ collapse onto a single universal curve 
$\tilde{y}(\tilde{x})$ independent of $\{p\}$. Note, that away from 
the Kondo scaling limit, this Kondo scale \TKscx can depend on the 
specific $x\in \{\omega,T,B,\ldots\}$ chosen. Moreover, the vertical 
normalization $y_0$ of the curves is not necessarily related to 
$\TKscx$. Rather, it depends on the measured quantity which may not 
even have units of energy. Typically, the specific choice for $y_0$ 
emerges out of context in a straightforward way, and as such is 
specified with each application below. 

\begin{table*}[tbh!]
  \begin{tabular}{l|lll|l|l} \hline\hline
     dependence on &
     \multicolumn{3}{l|}{universal Kondo scale $\TK=\tfrac{1}{4\chi_0}$}
     & correction to $\chid_0$ & see also \\ \hline
     $\omega$ or $T$ & \TKsc & where
     $\chisc_0$  & $=2 \chifs_0-\chid_0$ & $2 \times (\chifs_0-\chid_0)$ &
     \Eq{def:chi0}
  \\
     $B$ & \TKscB & where
     $\chiscB_0$ & $=\chifs_0$  & $1\times (\chifs_0-\chid_0)$ &
     \Eq{chi:fs}
  \\ \hline\hline
  \end{tabular}
\caption{Proposed corrections to the Kondo temperature
  based on the commonly used zero-temperature impurity
  susceptibility $\chid_0$ away from the strict scaling 
  limit of infinite bandwidth, yet in the Kondo regime
  having $\TK \lesssim 10^{-2}D$. In the scaling limit,
  all corrections vanish, \ie $\chifs_0 = \chid_0$.
}%
\label{tbl:TKs}
\end{table*}

The main result of this paper is the proposition of the altered
definition of the impurity susceptibility,
\begin{subequations}
\label{def:chi0}
\begin{align}
    \chisc(T)& \equiv
    \SPTn{\Sztot}{\Sztot} - \SPTn{\Szbath}{\Szbath} \label{def:chi0:a} \\
 &= 2 \SPTn{\Szimp}{\Sztot} - \SPTn{\Szimp}{\Szimp}
\text{,}  \label{def:chi0:b}
\end{align}
\end{subequations}
used for the scaling of dynamical or thermal quantities, \ie $x \in 
\{\omega, T\}$. Here $\Szbath \equiv \Sztot - \Szimp$ and $\langle 
\hat{S}^{\alpha} \Vert \hat{S}^{\beta}\rangle$ as defined with 
\Eq{def:chi0:1}. As will be demonstrated numerically, the definition 
of the susceptibility in \Eq{def:chi0} provides a sensitive Kondo 
scale through \Eq{def:TK:chi0}, \ie $\TKsc \equiv \lim_{T\to0} 
1/(4\chisc(T)) \equiv 1/(4\chisc_0)$, which allows for proper scaling 
(sc) of frequency or temperature dependent curves onto a single 
universal curve in a wide range of impurity parameters with bare 
energies from much smaller to much larger than the bandwidth $D$, 
provided that one has a well-defined Kondo regime, \ie $\TK \ll D$. 
For notational simplicity, $x$ will not be specified
with \TK here, \ie $\TKsc \equiv {\TKsc}^{,\omega} \equiv
{\TKsc}^{,T}$
A motivation of \Eq{def:chi0} in terms of the non-interacting system
is given in the \App{app:motivation}. More generally, as pointed out
with \App{app:motiv:nonint}, above scale-preserving susceptibility 
may be understood in terms of the scaling of frequency by the
quasi-particle weight $z$. \cite{Shastry13}

In contrast, the earlier definitions in \Eqs{def:chi0:1} can be 
reliably used for scaling in certain parameter regimes only (\eg the 
scaling limit when the bandwidth is the largest energy scale by far). 
The major differences of the impurity susceptibility in \Eq{def:chi0} 
to the definitions in \Eqs{def:chi0:1} are apparent. As compared to 
\Eq{def:chi0:Sztot}, the last term in \Eq{def:chi0:a} is calculated 
\textit{in the} \textit{presence} of the impurity. This comes with 
the benefit that, similar to \Eq{def:chi0:Szimp}, \Eq{def:chi0:b} can 
be computed entirely through the \textit{non-extensive} quantities 
since the extensive leading term in \Eq{def:chi0:a} cancels. 
Therefore, in contrast to \Eq{def:chi0:Sztot}, the impurity 
susceptibility in \Eq{def:chi0} can be computed for a given system 
without having to resort to a reference system without the impurity. 
Compared to \Eq{def:chi0:Szimp}, on the other hand, \Eq{def:chi0} 
acquires the relevant correction $\langle \Szimp \Vert \Szimp \rangle 
_{T}\rightarrow \langle \Szimp \Vert \Szimp \rangle _{T} - 2 [ 
\langle \Szimp \Vert \Szimp \rangle _{T} - \langle \Szimp \Vert 
\Sztot \rangle _{T} ] $. 

For the \TKsc derived from \Eq{def:chi0}, the emphasis is on a given
fixed Hamiltonian with infinitesimal perturbations whose (many-body)
excitations are explored either dynamically or thermally. For this,
the Kondo scale derived from $\chisc_0$ \emph{mimics} the scaling 
limit, even if the parameters that enter the Hamiltonian do not
strictly adhere to the scaling limit. In contrast, as will be shown
below, \emph{if the Hamiltonian itself} is altered through an
external parameter $x \in \{B, \ldots \}$ via $\hat{H}'=-x \hat{X}$,
universal scaling \vs a finite range in $x$ analyzed at zero
temperature is generally governed by a slightly different Kondo
scale, $\TKscx$, based on a variant of the impurity susceptibility
(henceforth, the notation \TKscx will be reserved for this context
only). 

In the scaling limit where bandwidth is the largest energy scale by 
far, it is found that $\langle \Szimp \Vert \Sztot \rangle _{T}\simeq 
\langle \Szimp \Vert \Szimp \rangle _{T}$ (for a proof of this in the 
non-interacting case, see \App{app:motiv:nonint}). Only in this 
regime, the static magnetic susceptibility can be computed 
equivalently in various ways including \Eqs{def:chi0:1}, \ie 
$\chisc(T) \simeq \chid(T) \simeq \chifs(T)$. Here, in particular, 
the more conventional magnetic susceptibility $\chid(T) $ may be 
replaced by $\chifs(T)$ which is much simpler and cheaper to 
evaluate.

The definitions for proper scale-preserving Kondo temperatures at 
finite bandwidth as proposed in this paper are summarized in 
\Tbl{tbl:TKs}. This includes the Kondo temperature \TKsc for fixed 
Hamiltonian for scaling of dynamical or thermal quantities, as well 
as the Kondo temperature \TKscB for scaling \vs an external parameter 
that alter the Hamiltonian at $T=\omega=0$, here for the specific 
case of magnetic field $B$. The derivation of the latter (see 
\Sec{sec:scalingB}) may also serve as a general guide for scaling \vs 
other external physical parameters that directly enter the 
Hamiltonian.

The remainder of the paper then is organized as follows: The rest of 
the introduction discusses the role of the new susceptibility 
$\langle \Szimp \Vert \Sztot \rangle _{T}$ introduced with 
\Eq{def:chi0} in terms of the Friedel sum rule (\Sec{subsec:FSR}). 
Furthermore, \Sec{sec:intro} still provides general computational 
aspects on the static linear susceptibility (\Sec{subsec:linsub}), 
followed by model conventions and methods (\Sec{subsec:models}). 
\Sec{sec:results} presents the results and discussion on the scaling 
of dynamical impurity spin susceptibility (\vs frequency), as well as 
the scaling of the linear conductance (\vs temperature and magnetic 
field). Following summary and outlook, the appendices provides 
detailed technical discussions. It includes (\App{app:motivation}) a 
motivation for the scale-preserving susceptibility which is mainly 
based on the non-interacting system, (\App{app:finitesize}) a 
technical discussion of finite-size effects of the dynamical impurity 
susceptibility, and (\App{app:chifs:NRG}) technicalities on the 
evaluation of the mixed susceptibility $\chifs(T)$ within the fdm-NRG 
framework. The latter also contains a short discussion on the 
evaluation of the impurity specific heat which, in a wider sense, 
also resembles the structure of an impurity susceptibility. Finally, 
\App{app:phaseshifts} comments on the conventional extraction of 
phase shifts from the many-body fixed-point spectra of the NRG, while 
also providing a detailed analysis of discretization, \ie finite 
size, effects.

\subsection{Magnetic susceptibility and Friedel sum rule
\label{subsec:FSR}}

The definition of the impurity susceptibility in
\Eq{def:chi0} introduces the additional impurity susceptibility,
\begin{equation}
   \chifs(T) \equiv \langle \Szimp \Vert \Sztot \rangle _{T}
 = \beta \langle \Sztot \Szimp \rangle _{T}
\text{,} \label{chi:fs}
\end{equation}
where $\beta\equiv 1/T$, and `$\mathrm{FS}$' stands for Friedel sum 
rule as motivated shortly. It will also be referred to as 
\emph{mixed} susceptibility, as it combines the impurity spin with 
the total spin. Assuming $B=0$, the last equality in \Eq{chi:fs} used 
$\langle \Sztot \rangle _{T} = \langle \Szimp \rangle _{T} = 0$. 
Given that $\Sztot$ commutes with the Hamiltonian, this reduces to 
the simple thermal expectation value as indicated, which can be 
evaluated efficiently (see \App{app:chifs:NRG} for details). 
Consequently, for $T=0^+$, this corresponds to a \emph{strict} 
low-energy quantity that that does not further explore the dynamics 
at intermediate or large frequency $\omega > T_\mathrm{K}$ [which is 
the case, for example, for the definition of the impurity 
susceptibility in \Eq{def:chi0:Szimp}]. 

The susceptibility in \Eq{chi:fs} can be interpreted twofold: (i) as
the local contribution to the total magnetization due to a global
external field, or equivalently, (ii) as the response in the total
magnetization of the system due to a local magnetic field at the 
impurity only. The first can be seen as (yet another) intuitive and
qualitative description of the local spin susceptibility. The latter
interpretation, on the other hand, allows a direct link to the
Friedel-sum-rule (FS) [hence the label in \Eq{chi:fs}]: given an
(infinitesimal) local change of the Hamiltonian. FS relates the
low-energy phase shifts $\varphi_\sigma$ of the entire system to the
\textit{total} change in local charge that flows to or from infinity
(note that this change in local charge includes the displaced charge
of both, the impurity itself as well as the close vicinity of the
impurity, which in total may simply be interpreted as displaced
``local'' charge \cite{Muender12}).

The dependence of the low-energy phase shifts $\varphi_\sigma$ of the 
bath electrons on an external magnetic field at the impurity can be 
used to define a Kondo scale $\TKphi$,\cite{Nozieres74} 
\begin{equation}
    \lim_{B\to 0}\tfrac{d}{dB} \varphi_\sigma
 \equiv \sigma \frac{\pi}{4\TKphi}
\ \text{,}\label{def:TKphi:0}
\end{equation}
evaluated at $T=0$, where $\sigma \in \{\uparrow,\downarrow \} \equiv 
\pm 1$. As a direct consequence of the Friedel-sum-rule then, it 
follows 
\begin{equation}
  \TKphi=\TKfs \text{,} \qquad (T=0)
\label{def:TKphi}
\end{equation}
since $\langle \Sztot \rangle = \tfrac{1}{2} ( \Delta N_{\uparrow} - 
\Delta N_{\downarrow} ) \overset{\mathrm{FS}}{=} \tfrac{1}{2\pi} 
\bigl( \varphi_{\uparrow} - \varphi_{\downarrow}\bigr) $, with 
$\Delta N_\sigma$ the change in total number of particles with spin 
$\sigma$ relative to $B=0$. Consequently, $\chifs \equiv 
\tfrac{d}{dB_\mathrm{imp}} \langle \Sztot \rangle = 1/(4\TKphi)$, 
which coincides with the definition of \TKfs, and hence proves 
\Eq{def:TKphi}. The identity in \Eq{def:TKphi} has also been verified 
numerically to within 1\% accuracy (using NRG with $\Lambda=2$ as 
defined below; for a more detailed discussion on the explicit 
extraction of phase shifts within the NRG, see 
\App{app:phaseshifts}). 

While, intuitively, one may have expected that the dependence of the 
low-energy phase shifts on the magnetic field yields a universal 
Kondo scale, this is true only in the specific case that data is 
scaled \vs magnetic field at $T=\omega=0$, \ie having $x=B$ (see 
\Sec{sec:scalingB} further below). However, this alters the 
Hamiltonian. For dynamical or thermal quantities for a given fixed 
Hamiltonian, having \Eq{def:chi0:b}, $\TKfs$ does not guarantee 
universal scaling. The reason for this may be seen as follows: while, 
in fact, the phase shifts themselves are not necessarily affected by 
finite bandwidth at $B=0^+$, \ie at the low-energy fixed point [\cf 
the discussion of $\chifs_0$ for the non-interacting case in 
\App{app:motiv:nonint}], when investigating an entire universal curve 
\wrt to frequency or temperature, this necessarily also explores 
states at intermediate energies. By exploring a range of energies, 
however, this becomes susceptible to finite bandwidth. Hence $\TKphi$ 
fails to provide proper scaling onto a universal curve for dynamical 
or thermal data. 

\subsection{Static linear susceptibility
\label{subsec:linsub}}

Consider the general static linear susceptibility for obtaining a
response in the measured operator $\langle \hat{Y}\rangle $ by
applying the infinitesimal external perturbation $\hat{H}'(\lambda )
=-\lambda \hat{X}$ to a given Hamiltonian,
\begin{equation}
   \SPTh{X}{Y} \equiv \lim_{\lambda \rightarrow 0} \tfrac{d}{d\lambda }
   \langle \hat{Y}\rangle _{T,\lambda }
 = \int\limits_{0}^{\beta }d\tau \,\langle\delta \hat{X}(\tau)
   \cdot \delta \hat{Y}\rangle _{T}
\text{,}\label{XY:suscept:1}
\end{equation}
with $\beta \equiv 1/T$, $\delta \hat{X}\equiv \hat{X}-\langle 
\hat{X} \rangle _{T}$, similarly for $\delta \hat{Y}$, and $\hat{X}( 
\tau ) \equiv e^{\tau \hat{H}}\hat{X}e^{-\tau \hat{H}}$ evaluated at 
$\lambda =0$. By definition, the operators $\hat{X}$ and $\hat{Y}$ 
are assumed hermitian. The last equality in \Eq{XY:suscept:1}, i.e. 
the imaginary-time Matsubara susceptibility, represents an 
\emph{exact} mathematical relation, \cite{LeBellac04} which satisfies 
the properties of a scalar product for hermitian operators, \ie 
$\SPTh{X}{Y} \equiv \SPTh{Y}{X}^{^\ast}$ with $\SPTh{X}{X} \ge 0$ 
(\cf Bogoliubov-Kubo-Mori scalar product [\onlinecite{LeBellac04}]). 
If $\hat{X}$ and $\hat{Y}$ do not commute with the Hamiltonian and 
$\langle \hat{X}\rangle _{T}=\langle \hat{Y}\rangle _{T}=0$, then 
\Eq{XY:suscept:1} is equivalent to the Kubo formula for linear 
response in the thermodynamic limit, 
\begin{equation}
   \SPTh{X}{Y} \simeq \SPTh[R]{X}{Y} \equiv
   -\lim_{\omega \rightarrow 0}\chiXY( \omega)
\label{XY:suscept:2}
\end{equation}
with $\chiXY( \omega ) $ the Fourier transformed dynamical retarded 
(R) correlation function $\chiXY( t) \equiv -i\vartheta (t) \langle 
\lbrack \hat{X}( t) ,\hat{Y} ]\rangle _{T}$ [the sign with the last 
term in \Eq{XY:suscept:2} originates in the sign of the definition of 
$\hat{H}'$ with \Eq{XY:suscept:1} which ensures a positive 
susceptibility for $\hat{X}=\hat{Y}$]. The Kubo formula as in 
\Eq{XY:suscept:2}, however, assumes that the system has no long-time 
memory of the applied operators $\hat{X}$ or $\hat{Y} $. Importantly, 
for exactly this reason for discretized, i.e. effectively finite-size 
systems, only \Eq{XY:suscept:1} represents a reliable working 
definition, whereas corrections can apply to \Eq{XY:suscept:2} [\eg 
see App.~\ref{app:finitesize}]. Most notably, if the Hamiltonian 
preserves total spin (which will be assumed throughout this paper), 
then with $\hat{X}=\hat{Y}=\Sztot $, the resulting dynamical 
correlation function $\mathrm{Im}\,\chi ( \omega ) \propto 0\cdot 
\delta ( \omega ) $ is pathological. In contrast, \Eq{XY:suscept:1} 
yields the correct result $\SPTn{\Sztot}{\Sztot} = \beta \langle 
(\Sztot)^{2}\rangle _{T} - \langle \Sztot \rangle _{T}^{2} \equiv 
\beta \,\Delta ^{2}\Sztotv$, i.e. the thermal fluctuations in the 
total spin of the system, using the grand-canonical ensemble in the 
evaluation of the thermal average $\langle \cdot \rangle _{T}$. 

\subsection{Models and method \label{subsec:models}}

A prototypical quantum impurity model is the single impurity Anderson 
model (SIAM). \cite{Friedel54,Anderson67} It consists of the local 
Hamiltonian, $\hat{H}_0^\mathrm{SIAM} \equiv \hat{H}_\mathrm{imp} + 
\hat{H}_\mathrm{cpl}$, with 
\begin{subequations}\label{eq:SIAM}
\begin{align}
   \Himp &=
      \sum_\sigma \varepsilon_{d\sigma} \hat{n}_{d\sigma}
      + U \hat{n}_{d\uparrow} \hat{n}_{d\downarrow}
\label{eq:SIAM:Himp} \\
   \Hcpl &= \sum_{k\sigma} \bigl(
      V_{k\sigma} \hat{d}_{\sigma}^{\dagger}
      \hat{c}_{k \sigma}^{\phantom{\dagger}} + \Hc \bigr)
    \equiv  \sqrt{\tfrac{2D\Gamma}{\pi}} \sum_\sigma (
       \hat{d}_{\sigma}^{\dagger} \hat{f}_{0 \sigma}^{\phantom{\dagger}}
     + \Hc )
\text{.}\label{eq:SIAM:Hcpl}
\end{align}
\end{subequations}
It describes a single interacting fermionic (d-)level, \ie the 
impurity (imp), with level-position $\varepsilon_{d\sigma}$ and 
onsite interaction $U$, which is coupled (cpl) through hybridization 
to a non-interacting macroscopic Fermi sea $\Hbath \equiv 
\sum_{k\sigma} \hat{n}_{k\sigma}$ with $\varepsilon_{k\sigma} \in 
[-D,D]$ of half-bandwidth $D:=1$ (all energies taken in units of $D$, 
unless specified otherwise). Here $\hat{d}^\dagger_{\sigma}$ 
($\hat{c}^\dagger_{k\sigma}$) creates an electron with spin $\sigma 
\in \{\uparrow,\downarrow\}$ at the d-level (in the bath at momentum 
$k$), respectively, with $\hat{n}_{d\sigma} \equiv 
\hat{d}^\dagger_{\sigma} \hat{d}_{\sigma}$, and $\hat{n}_{k \sigma} 
\equiv \hat{c}_{k \sigma}^\dagger \hat{c}_{k 
\sigma}^{\phantom{\dagger}}$. If a magnetic field is applied at the 
impurity (in the bath), then $\varepsilon_{d\sigma}=\varepsilon_{d} - 
\tfrac{\sigma}{2}B$ ($\varepsilon_{k\sigma} = 
\varepsilon_{k}-\tfrac{\sigma}{2}B$), respectively. The sign has been 
chosen such, that for $B>0$ a positive magnetization 
$\langle\Sz\rangle$ arises. With $\nu$ the density of states, 
$\Gamma_{\sigma}(\varepsilon)\equiv\pi\nu V^2_\sigma(\varepsilon) = 
\Gamma\cdot\theta(D-|\omega|)$ is the hybridization strength. It is 
taken constant and the same for each spin $\sigma$, for simplicity. 

In the limit of large $U$, the SIAM reduces to the Kondo model with a 
singly occupied impurity (a fluctuating spin), which couples to the 
electrons in the bath through the spin-spin interaction 
\cite{Kondo64,Hewson93} 
\begin{align}
   \hat{H}_0^\mathrm{Kondo} &= 2J\,
   \mathbf{\hat{S}}_d \cdot \mathbf{\hat{S}}_0
\end{align}
with $J>0$ the antiferromagnetic Heisenberg coupling (using constant
density of states $\nu=1/2D$ of the bath, for simplicity), 
\cite{Hewson93} $\hat{S}_d$ the spin operator of the impurity and
$\hat{S}_0^x \equiv \tfrac{1}{2} \sum_{\sigma\sigma'}
\hat{f}_{0\sigma}^{\dagger} \tau^{x}_{\sigma,\sigma'} \hat{f}_{0
\sigma'}^{\phantom{\dagger}}$ the normalized spin operator of the
bath site $\hat{f}_{0\sigma}$ at the location of the impurity with 
$\tau^{x}$ the Pauli spin matrices ($x\to\{x,y,z\}$).

The generic interacting impurity setting above involves the solution 
of a strongly-correlated quantum many-body system, which can be 
simulated efficiently using the quasi-exact numerical renormalization 
group (NRG). \cite{Wilson75, Bulla08} In order to deal with arbitrary 
temperatures in an accurate manner, the fdm-NRG is employed 
\cite{Wb07,Wb12_SUN, Wb12_FDM} which is based on complete basis 
sets.\cite{Anders05} While not explained in detail here (for this see 
Refs.~[\onlinecite{Wilson75,Bulla08,Wb12_FDM}]), the essential NRG 
related computational parameters indicated with the figures below are 
the dimensionless logarithmic discretization parameter 
$\Lambda\gtrsim 2$, the truncation energy \Etrunc in rescaled units 
(as defined in [\onlinecite{Wb12_FDM}]), the number $N_z$ of 
$z$-shifts for $z$-averaging, \cite{Oliveira91} and the log-Gaussian 
broadening parameter $\sigma$ for smooth spectral data.

\begin{figure}[tb!]
\begin{center}
\includegraphics[width=1\linewidth]{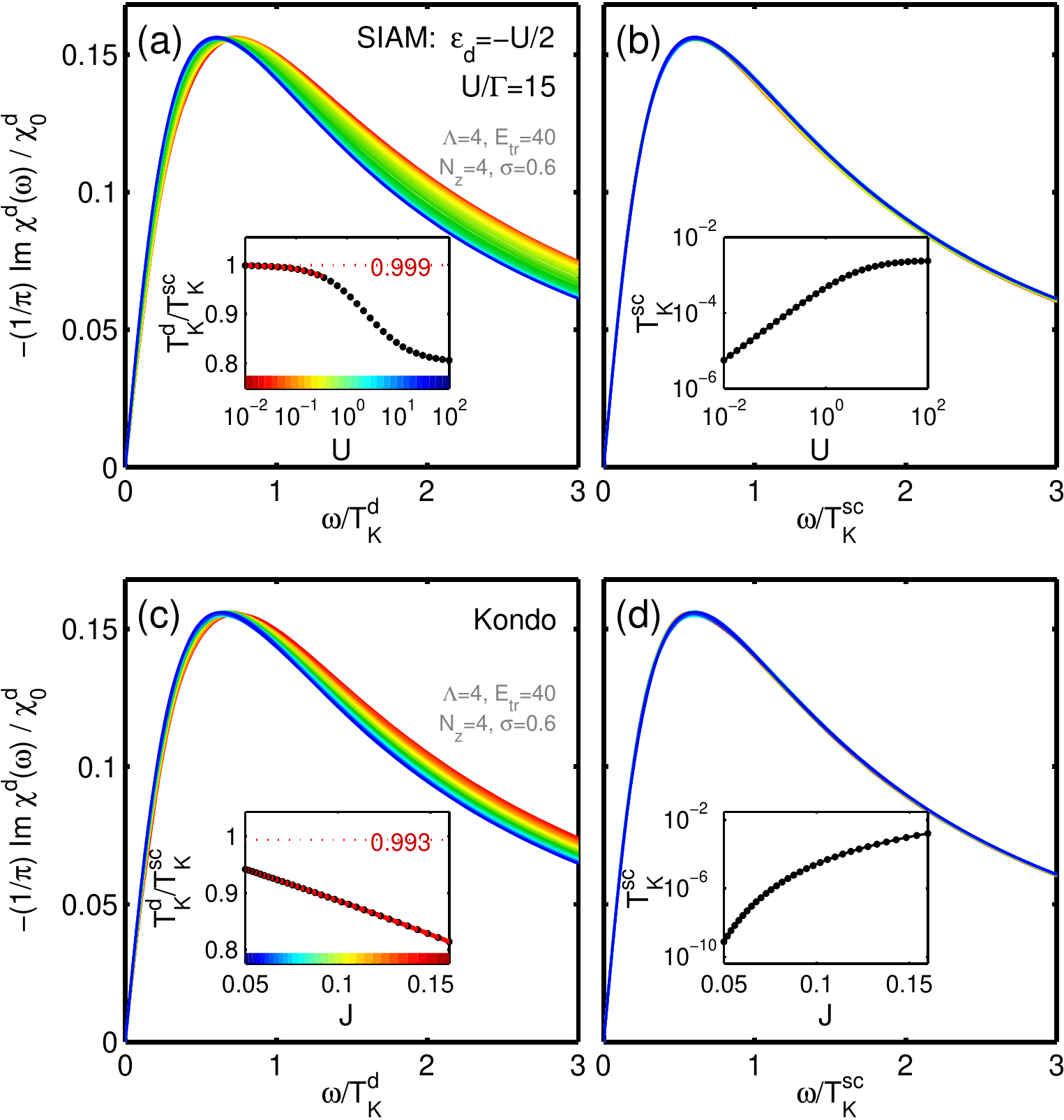}
\end{center}
\caption{(Color online)
   Scaling of the frequency of the dynamical spin susceptibility
   $\chid(\omega) / \chid_0$ by the conventional impurity susceptibility
   $\TKd \equiv 1/(4\chid_0)$
   (left panels) vs. the scale-preserving definition of Kondo
   temperature $\TKsc \equiv 1/(4\chisc_0)$ (right panels): all the
   densely lying curves of the left panels collapse onto a single
   universal curve in the right panels, respectively.
   The upper panels (a and b) analyze the SIAM. The inset to panel (a)
   demonstrates the dependence of $\TKd/\TKsc$ vs. the onsite
   interaction $U$, 
   while keeping the
   ratios $U/\Gamma=15$ and $\varepsilon_d =-U/2$ fixed. The color bar
   at the bottom of the inset relates the color of the lines in the
   main panel to the specific values of $U$ ranging from $U\ll 1$ to $U
   \gg 1$ (with $D\equiv1$ the bandwidth).
   The limit $\lim_{U\to0} [\TKd/\TKsc]$ has been fitted, resulting
   in the value of 1. with excellent accuracy (actual
   value indicated together with the horizontal dotted line).
   The inset to panel (b), shows the dependence of \TKsc vs. $U$
   which stretches over several orders of magnitude.
   In complete analogy, the lower panels (c and d) analyze the Kondo
   model. In particular, the fitted limit $\lim_{J\to0} \TKd/\TKsc \simeq 1$ in
   the inset of panel (c) is the same as for the SIAM (cf. panel a)
   within the numerical error of significantly less than 1\%
   [for comparison, the same calculation yet with the cheaper and less
   accurate setting of $\Lambda=2$ and $\Etrunc=12$ (not shown)
   already resulted in $\TKd/\TKsc\simeq 0.98$, while
   $\Lambda=4$ and $\Etrunc=20$ (not shown) already agreed well
   with above results. In this sense, above results for
    $\Lambda=4$ and $\Etrunc=40$ are considered fully converged].
} \label{fig:chi:dyn}
\end{figure}

\section{Results and Discussion \label{sec:results}}

\subsection{Scaling of dynamical susceptibility}

The dynamical magnetic susceptibility of the impurity is analyzed in 
\Fig{fig:chi:dyn} for both the SIAM (upper panels) as well as the 
Kondo model (lower panels) for a wide range of parameters, resulting 
in a dense set of curves. For the left panels, the horizontal 
frequency axis is scaled by $\TKd \equiv 1/(4\chi_0^\mathrm{d})$, 
which clearly fails to reproduce a single universal curve. The 
universal scaling is provided only by the scaling of frequency using 
the altered \TKsc (right panels). The residual tiny deviations stem 
from the data with largest \TK, \ie with $T_\mathrm{K} \gtrsim 
10^{-3} D$. 

By analyzing the universal scaling at an accuracy of $\lesssim 1\%$, 
this required at the very minimum a parameter setting in the strongly 
correlated Kondo regime. Hence the Kondo temperature was kept clearly 
smaller than the bandwidth, \ie $\TK < 10^{-2}$. 
For the SIAM, this allowed a wide range for the interaction strength 
from significantly smaller to significantly larger than the 
bandwidth,\cite{Hofstetter99} nevertheless, while keeping 
$\Gamma/U=\tfrac{1}{15}$ and $\varepsilon_d/U=-\tfrac{1}{2}$ constant 
[\cf \Fig{fig:chi:dyn}(a); similarly, the scaling was also tested 
away from the particle-hole symmetric point at $\varepsilon_d/U = 
-\tfrac{1}{3}$, resulting in equally excellent scaling of the data 
(not shown). The scaling also was tested for the non-interacting case 
($U=\varepsilon_d=0$ yet finite $\Gamma$; not shown) where $\Gamma$ 
takes the role of \TK. As a consequence, in complete analogy to 
above, for $\Gamma < 10^{-2}$ this allowed for similar excellent 
scaling of the data, yet, of course, to a different universal curve]. 

The different definitions of the Kondo temperature, $\TKd$ vs. 
$\TKsc$, are analyzed in the insets of the left panels, showing clear 
deviations of \TKsc from \TKd of up to 20\%, with \TKd consistently 
smaller than \TKsc. The deviations are more pronounced for the Kondo 
model, remembering that this essentially reflects the large-$U$ limit 
of the Anderson model, which implies $U\gg D$ (even for Kondo 
temperatures as small as $\TK\simeq 10^{-10}$, the difference between 
$\TKd$ and $\TKsc$ is still about 6\% [see inset in lower panels]). 
In the limit $\TK \to 0$ both, the SIAM ($U\to 0$ with appropriately 
adjusted $\Gamma$ and $\varepsilon_d$) as well as the Kondo model 
($J\to 0$) result in the same ratio $\TKd/\TKsc = 1$ within the 
accuracy of the fitted extrapolations in the insets (using 
$3^\mathrm{rd}$ order polynomials with the fitting range indicated 
with the fit in red on top of the data; see caption on the 
convergence of $\TKd/\TKsc$ with varying NRG parameters).

\begin{figure*}[tbh!]
\begin{center}
\includegraphics[width=1\linewidth]{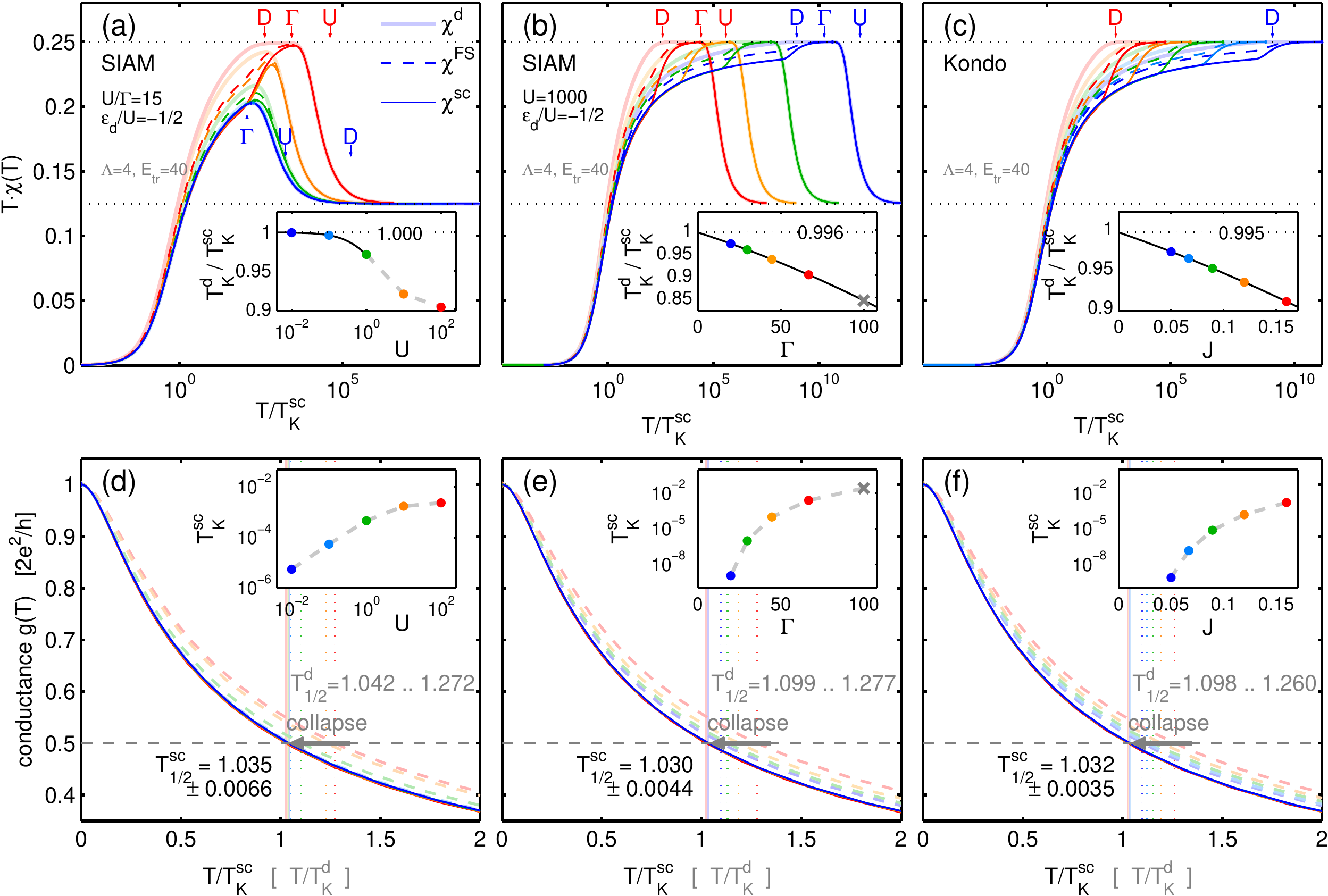}
\end{center}
\caption{(Color online)
   Temperature dependent scaling of the static spin susceptibility
   $\chi(T)$ (upper panels) and the linear conductance $g(T)$
   (in units of $2e^2/h$; lower panels)
   for the SIAM (left and center panels),
   as well as for the Kondo model (right
   panels). The color of the lines in the main panels matches
   the colors of the symbols in the inset, hence this indicates
   the respective parameter setting.
   The upper panels compare various definitions of the static
   spin susceptibility ($\chid$, $\chifs$, $\chisc$ in faint,
   dashed and solid, respectively).
   In the upper main panels, for clarity, the actual value of 
   the relevant parameters [$\{D,\Gamma,U\}$ for panels (a-b)
   and $D$ for panel (c)] are indicated in units of \TKsc
   for the largest and smallest $\TK$ only.
   Similar to \Fig{fig:chi:dyn}, the insets to the upper panels
   analyze the relation between \TKd and \TKsc as function of the
   parameters. Their ratio is fitted towards $\TK\to 0$, resulting in a
   comparable value of $1$ to very good accuracy as indicated
   for all three cases (panel a-c). The actual exponential
   range of \TKsc is shown in the insets to the lower panels.
   The lower panels show the static linear conductance $g(T)$ \vs
   $T/\TKd$ (non-universal; dashed faint lines, but color
   match with symbols of inset otherwise) and \vs $T/\TKsc$ (solid
   lines) which show proper scaling behavior, in that all lines
   collapse onto a single universal curve.
   With \THg the temperature where $g(T)$ passes through
   $1/2$, in units of \TKd, this ranges from $\THd \equiv
   T_{\nicefrac{1}{2}} / T_\mathrm{K}^\mathrm{d} =1.25$
   down to 1.03 [indicated by the vertical dotted lines
   with the range of \THd specified with each panel (gray text at center
   right in each panel)]. In units of \TKsc, this range collapses
   to the fixed value of  $T^\mathrm{sc}_{\nicefrac{1}{2}}
   \equiv T_{\nicefrac{1}{2}} / T_\mathrm{K}^\mathrm{sc} \simeq
   1.03 $ to within residual relative
   variations of clearly less than 1\%
   for all three cases [panels d-f; indicated by vertical
   solid light lines with their range specified by \THsc (black text)].
   Using $\Lambda=4$ and $\Etrunc=40$ as indicated,
   the value of $\THsc\simeq 1.03$ above is considered well
   converged [for comparison, for $\Lambda=2$ and $\Etrunc=8$
   a similar calculation (not shown) resulted in $\THsc\simeq 0.99$,
   while  $\Lambda=2$ and $\Etrunc=12$ resulted in
   $\THsc\simeq 1.01$; while good overall scaling can already
   be observed for $\Etrunc\lesssim 10$, the minor
   variations for smaller \Etrunc can be mostly eliminated
   by normalizing $g(T)$ by the numerical value $g(0)\approx
   1$ which was not included here].
}
\label{fig:chi:gg}
\end{figure*}

\begin{figure*}[tb!]
\begin{center}
\includegraphics[width=1\linewidth]{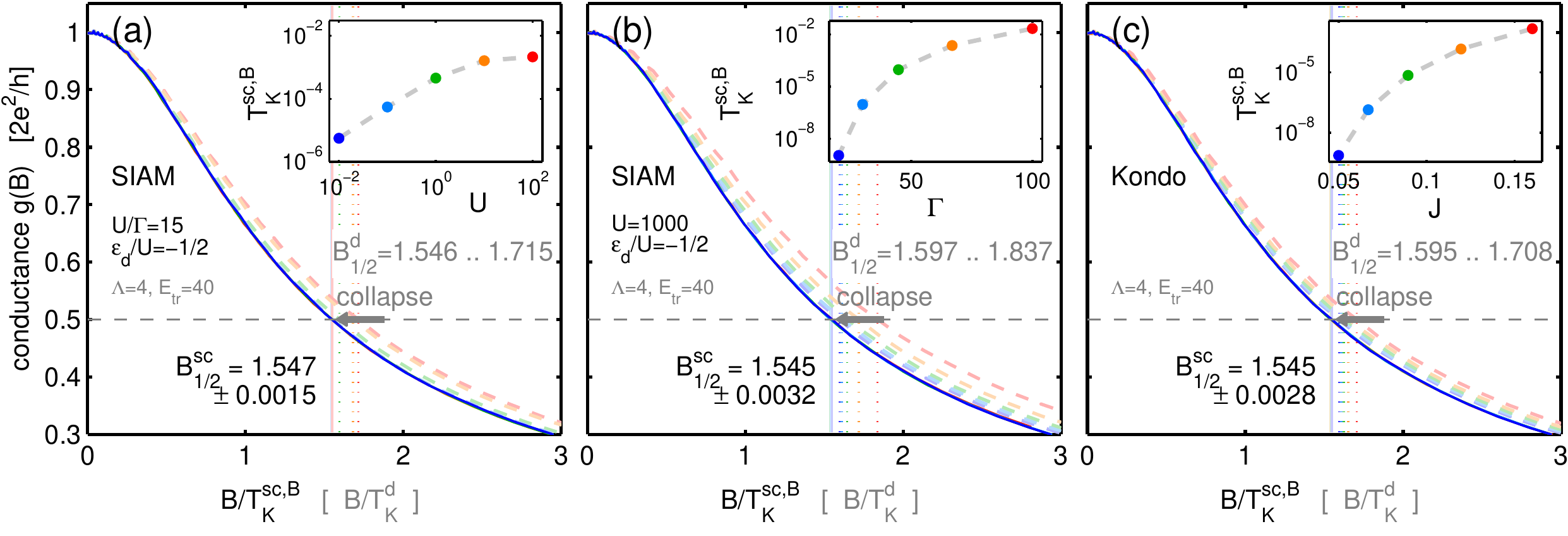}
\end{center}
\caption{(Color online)
   Linear conductance \vs magnetic field at $T=0$
   for the SIAM (left and center panel), as well as for 
   the Kondo model (right panel).
   Again the insets indicate the
   respective parameter setting of the lines in the main panels.
   Analogous to the analysis in \Fig{fig:chi:gg}(d-f),
   here the main panels show the static linear conductance $g(B)$ \vs
   $B/\TKd$ (non-universal; dashed faint lines, but color
   match with symbols of inset otherwise) and \vs $B/\TKscB$ (solid
   lines) which demonstrate universal scaling.
   With \BHg the magnetic field where $g(B)$ passes through $1/2$,
   in units of \TKd this changes from $\BHd \equiv
   B_{\nicefrac{1}{2}} / T_\mathrm{K}^\mathrm{d} =1.84$
   down to 1.55 for given data [indicated by the vertical dotted
   lines with their individual range specified with each panel
   (gray text at center
   right in each panel)]. In units of \TKscB this range
   collapses to the value $\BHsc \equiv B_{\nicefrac{1}{2}} /
   T_\mathrm{K}^\mathrm{sc} =1.55$ to within relative uncertainties
   of clearly less than 1\%
   for all three cases [panels d-f; indicated by vertical solid
   light lines with the range \THsc specified by the black text].
   Using $\Lambda=4$ and $\Etrunc=40$ as indicated, the data is
   considered fully converged (regarding minor variations for
   significantly lower $\Etrunc\lesssim 10$ and thus much faster
   calculations, see caption to \Fig{fig:chi:gg}).
} \label{fig:gB}
\end{figure*}

\subsection{Scaling of static susceptibility
and linear conductance \vs temperature ($B=0$)}

The scaling of the static magnetic susceptibility and the linear 
conductance of the SIAM and Kondo model \vs temperature is analyzed 
in \Fig{fig:chi:gg}. The left panels analyze the SIAM in a wide range 
of the onsite interaction $U$. The center panels analyze the SIAM 
still, yet in the large-$U$ limit while varying $\Gamma$, thus 
transitioning to the Kondo model. The right panels, finally, analyze 
the Kondo model itself. In all cases the parameters were chosen such 
that $\TK \lesssim 10^{-2}$ with \TK plotted in the insets with the 
lower panels (the \TK for the largest $\Gamma$ in the center panels 
exceeded $10^{-2}$ hence was excluded from the scaling analysis as 
indicated by the gray cross in the insets for the center panels). 

The quantity $T\cdot\chi(T)$ as plotted in the upper panels of
\Fig{fig:chi:gg} for the spin susceptibility, reflects
spin-fluctuations at the impurity. The high-temperature limit for
the Anderson (Kondo) impurity is given by $1/8$ ($1/4$),
respectively, indicated by the horizontal dashed lines. Clearly,
once $T$ exceeds $U$ for the SIAM (or $D$ for the Kondo model), the
large temperature limit is rapidly and accurately approached for
either definition of the impurity susceptibility. For the SIAM, for
$U\gg D$ an intermediate regime $D<T<U$ emerges which represents a
free spin, consistent with $T\cdot\chi(T) \to \tfrac{1}{4}$
[\Fig{fig:chi:gg}(a-b)]. For the Kondo model [\Fig{fig:chi:gg}(c)],
this regime is represented by $T>D$.

In the regime $U\ll D$ for the SIAM, the effective bandwidth relevant 
for the impurity is given by $U$, such that the actual full bandwidth 
$D$ of the Fermi sea becomes irrelevant in the description of the 
impurity [see $U=10^{-2}$ data (dark blue) in \Fig{fig:chi:gg}(a)]. 
As a consequence, here the impurity susceptibility is rather 
insensitive to its precise definition, \ie $\chid(T) \simeq \chifs(T) 
\simeq \chisc(T)$ [see $U=10^{-2}$ data in inset to 
\Fig{fig:chi:gg}(a)], which thus is considered a safe regime for 
local susceptibility calculations and subsequent Kondo scaling. The 
differences between the three definitions of the impurity 
susceptibility, however, become strongly visible as $U$ increases and 
surpasses the bandwidth [\eg see $U=10^{2}$ data (red curves) in 
\Fig{fig:chi:gg}(a)]. This behavior is precisely also reflected in 
the zero-temperature ratios $\TKd/\TKsc$ as shown in the inset to 
\Fig{fig:chi:gg}(a), which strongly deviate from $\approx 1$ as $U$ 
increases. 

For fixed large $U\gg D$, \TK can be strongly varied by tuning the 
hybridization $\Gamma$. The resulting data for the magnetic 
susceptibility is shown in \Fig{fig:chi:gg}(b). By plotting 
temperature in units of \TKsc, the data for $\chisc(T)$ nicely 
collapses onto a universal curve for $T<D$, a feat which, in 
particular, cannot be achieved for $\chid(T)$ in a similarly accurate 
manner. Furthermore, having $U\gg D$, the data in \Fig{fig:chi:gg}(b) 
for $T<U$ clearly resembles the Kondo model, as can be seen by direct 
comparison to the data of the actual Kondo model in 
\Fig{fig:chi:gg}(c). 

The lower panels of \Fig{fig:chi:gg} analyze the scaling of the 
linear conductance as measured in transport through a quantum dot 
which represents a prototypical \emph{quantum impurity} 
setting.\cite{GoldhaberGordon98,Kretinin11} It is computed by folding 
the impurity spectral function $A_\sigma(\omega;T) \equiv \int 
\tfrac{dt}{2\pi} \langle \{\hat{d}_\sigma(t),\hat{d}_\sigma^\dagger 
\}\rangle_T$ with the derivative of the Fermi distribution function, 
\ie $g(T)=\tfrac{\pi\Gamma}{2} \sum_\sigma \int\,d\omega\, 
A_\sigma(\omega;T)(-\tfrac{df}{d\omega})$ in units of $2e^2/h$. When 
scaling the temperature by \TKd, the resulting data is plotted in 
light dashed lines, which show a clear non-universal spread akin to 
the earlier analysis in \Fig{fig:chi:dyn}(a). In particular, the 
temperature $\THd$ where $g(T)$ passes through $1/2$ changes from 
1.25 down to 1.03 in units of \TKd, with the large-$U$ regime for the 
SIAM [\Fig{fig:chi:gg}(e)] and in particular also the Kondo model 
itself [\Fig{fig:chi:gg}(f)] most strongly affected. In contrast, 
when scaling the temperature by \TKsc, again an excellent scaling 
collapse is observed (solid lines in lower panels of 
\Fig{fig:chi:gg}). Note, furthermore, that the resulting $\THsc 
\equiv \THg/\TKsc = 1.032\pm 0.005$ nicely agrees across all panels 
from the SIAM [\Fig{fig:chi:gg}(a-b)] to the Kondo model 
[\Fig{fig:chi:gg}(c)], despite the broad parameter range analyzed. 
Given $\Lambda=4$ together with $\Etrunc=40$, these results are 
considered well converged [see figure caption on the convergence of 
$\THg/\TKsc$ with NRG parameters]. Finally, note that the value for 
$\THg/\TKsc$ above also agrees well with the one cited by Merker et 
al.\cite{Merker13} which in the wide-band limit suggests $\THg/\TKsc 
\simeq 1.04$. 
Overall, with $\THg/\TKsc$ being constant, this is fully consistent 
with the fact that $\THg$ itself may serve and is frequently used as
a universal definition of \TK, with a minor constant proportionality
factor of $1.03$ to the \TKsc used here.

Above results have direct implications on the Fermi liquid 
coefficients derived from the conductance $g(T)$. For example, with 
the Fermi liquid coefficient $c_T$ defined by $g(T) \simeq 1-c_T 
(T/\TK)^2$ for $T\ll \TK$, \cite{Nozieres74, Grobis08, Pletyukhov12, 
Merker13} this strongly depends on the precise definition of \TK. 
Note that even though \TK is apparently well-defined through the 
magnetic susceptibility, depending on the precise definition of the 
latter, nevertheless variations of up to 10\% are seen in the ratio 
$\TKd/\TKsc$ within a well-defined Kondo regime [\cf insets to upper 
panels of \Fig{fig:chi:gg}]. Therefore when using \TKd, this 
systematically underestimates $c_T$ by up to 20\%. It follows from 
the present analysis that the correct choice for \TK in the 
definition of $c_T$ is \TKsc, as it reflects the scaling limit, 
despite using parameters that do not strictly represent the scaling 
limit itself. Note, however, that the strict scaling limit is given 
by the regime $\TKd/\TKsc \simeq 1$, which for the Kondo model 
through the inset to \Fig{fig:chi:gg}(c) implies $J\lesssim 0.01$, 
resulting in the extremely small and rather impractical $\TK \lesssim 
10^{-45}$.

\subsection{Scaling of linear conductance \vs magnetic field ($T=0$)
\label{sec:scalingB}}

The linear conductance at finite magnetic field yet zero temperature 
is a strict low-energy quantity, in that $g(B) = \tfrac{\pi\Gamma}{2} 
\sum_\sigma A_\sigma (\omega=0;B,T=0)$ requires the spectral function 
evaluated at $\omega=0$ only. As a consequence, its sensitivity on 
finite bandwidth is minimal (\cf \App{app:motivation}). This already 
suggests that in given case where the Hamiltonian is \emph{altered} 
by a finite external parameter, universal scaling is not governed by 
the same \TKsc as introduced in \Eq{def:chi0}. 
Instead, through the Landauer formula, which in given case implies 
$\pi\Gamma\cdot A_\sigma(\omega=0;B,T=0) = \sin^2( \varphi_\sigma(B) 
)$, the conductance can be directly linked to the spin-dependent 
low-energy phase shifts $\varphi_\sigma$ of the entire system. For 
given particle-hole symmetric case, these can be written as 
$\varphi_\sigma(B) = \tfrac{\pi}{2} + \delta_\sigma(B)$ where for 
$|B| \ll \TK^\mathrm{(FS)}$, $\delta_\sigma(B) \equiv \sigma \pi B 
/(4\TKfs)$ [\cf \Eq{def:TKphi:0}] with $\sigma \in 
\{\uparrow,\downarrow\} \equiv \pm 1$. This directly identifies \TKfs 
as defined in \Eq{chi:fs} as the relevant Kondo temperature for 
universal scaling. Specifically, one obtains, 
\begin{align}
g(B) & 
      = \tfrac{1}{2}\sum_\sigma \sin^2(\varphi_\sigma)
 \simeq \tfrac{1}{2}
        \sum_\sigma (1 - \tfrac{1}{2} \delta_\sigma^2)^2  \notag\\
&\simeq 1- \bigl(\tfrac{\pi B}{4\TKfs}\bigr)^2
  \equiv 1- c_T \bigl(\tfrac{B}{\pi \TKfs}\bigr)^2
\label{gB:FL:scaling}
\end{align}%
with $c_T \equiv \tfrac{\pi^4}{16}$
the well-known Fermi-liquid coefficient \wrt temperature
for Kondo impurities.
\cite{Nozieres74,Grobis08,Pletyukhov12,Merker13}

The scaling of the linear conductance $g(B)$ with \TKfs is 
demonstrated in \Fig{fig:gB} for values of $B$ that stretch well 
beyond the quadratic regime in \Eq{gB:FL:scaling}. The analysis in 
\Fig{fig:gB} is completely analogous to \Fig{fig:chi:gg}(d-f), except 
that here the dependence is on the magnetic field. Consistent with 
the earlier analysis, the data for the SIAM with smallest $U=0.01$ in 
\Fig{fig:gB}(a) already closely resembles the scaling limit. In 
contrast, the curves for the Kondo model in \Fig{fig:gB}(c) even for 
the smallest coupling $J$ with its extremely small \TK still do not 
strictly represent the scaling limit. 

Above scaling analysis for $g(B)$ has major consequences for the 
extraction of the Fermi-liquid coefficient $c_B$, defined by $g(B) 
\simeq 1-c_B (B/\TK)^2$ for $B\ll \TK$ at $T=0$. 
\cite{Nozieres74,Grobis08,Pletyukhov12,Merker13} Above analysis 
suggests that the Kondo scale, that needs to be considered for an 
accurate evaluation of $c_B$ in a practical setting, is \TKfs. This 
then again resembles the scaling limit while, nevertheless, it allows 
to use finite or narrow bandwidth in ones analysis provided that $\TK 
\lesssim 10^{-2}$ (in units of $D$ as always). 

\section{Summary and outlook \label{sec:summary}}

In summary an adapted scheme for the calculation of the local 
susceptibility has been introduced which, at zero temperature, allows 
to define a proper universal Kondo scale \TKsc. The latter fully 
respects scaling of measured low-energy properties such as Kondo 
related features. A distinction needs to be made between dynamical or 
temperature dependent quantities which are described by the same 
fixed Hamiltonian (\TKsc), as compared to dependence on external 
parameters which directly enter the Hamiltonian, such as magnetic 
field (\TKscB). The corrections to the commonly used \TK based on the 
local susceptibility $\chid_0$ have been summarized in \Tbl{tbl:TKs}. 
For the parameter sets analyzed in this paper, these corrections 
range from about 0 to 10\% (which become about twice as large still 
for Fermi liquid coefficients), yet vanish in the scaling limit. 

The effect of finite bandwidth on the Kondo scale was discussed, 
while assuming a featureless hybridization otherwise. Proper scaling 
was demonstrated for the SIAM in a broad parameter regime, with the 
interaction $U$ ranging from much smaller to much larger than the 
bandwidth $D$. The latter large-$U$ limit then also was shown to 
smoothly connect the SIAM to the Kondo model. Essentially, this is 
the numerical equivalent of the Schrieffer-Wolff transformation 
without actually making any approximation. \cite{Wb12_SUN} By 
construction, the effects of finite bandwidth are clearly most 
prominent in the large-$U$ limit ($U\gg D$), and as a consequence 
also affect most strongly the Kondo model itself. The discussion of a 
universal low-energy scale for specific model parameters away from 
the abstract true Kondo scaling limit with the bandwidth by far the 
largest energy is important in the experimental context, but also in 
the numerical context by choosing a parameter regime where 
simulations can be performed more efficiently (\eg Kondo model \vs 
SIAM). The explicit analysis and discussion of the universal Kondo 
scale applied to Fermi-liquid coefficients is beyond the scope of 
this paper, and will be published elsewhere. 

Finally, it is pointed out that the impurity contribution to the 
specific heat, $c_V(T)$, essentially also has the structure of a 
susceptibility, namely the response in energy at the impurity due to 
an increase in the external parameter $T$, \ie the temperature. The 
analogies remain vague, though, since temperature is special as 
compared to other external parameters such as magnetic field as it 
enters in the Boltzmann distribution for thermal statistics. 
Moreover, it is also unclear a priori whether and to what extent to 
associate the coupling term $\Hcpl$ with the impurity or the bath. 
Nevertheless, an approximate expression for the impurity contribution 
to the specific heat can be evaluated by computing $c_V(T) \simeq 
\tfrac{d}{dT} \langle \Himp + \tfrac{1}{2}\Hcpl \rangle_T$. 
\cite{Merker12cv} In contrast to [\onlinecite{Merker12cv}], however, 
which computes $c_V(T)$ by the explicit numerical derivative \wrt 
temperature, the latter can be fully circumvented along the lines of 
the mixed susceptibility \chifs discussed above by directly computing 
the plain thermal expectation value $\beta \langle \Himp+\tfrac{1}{2} 
\Hcpl \Vert \Htot \rangle_T = \beta^2 \langle (\Himp+\tfrac{1}{2} 
\Hcpl) \Htot \rangle_T$ within the fdm-NRG framework [see 
\App{app:specific:heat} for details]. 

\begin{acknowledgements}

We want to thank Oleg Yevtushenko,  Herbert Wagner, and Jan von
Delft for fruitful discussions, and also Theo Costi and Mikhail
Pletyukhov for their comments on the manuscript. This work has
received support from
DFG (TR-12, SFB631, NIM, and WE4819/1-1).

\end{acknowledgements}

\appendix

\section{Motivation for scale preserving susceptibility at $T=0$
\label{app:motivation}}

The definition of the magnetic susceptibility $\chi^\imp(T)$ in 
Eq.~(\ref{def:TK:chi0}) is typically computed through its spectral 
function $\chi''(\omega) \equiv -\tfrac{1}{\pi} \mathrm{Im} 
\chi^\mathrm{R}(\omega)$, having $\chi(\omega) \equiv \chi'(\omega) - 
i\pi\chi''(\omega)$ [for simplicity, the following discussion only 
refers to the static local impurity susceptibility $\chi^\imp(T)$, 
hence the superscript $d$ will be skipped for readability]. This 
spectral function is given by 
\begin{align}
   \chi''(\omega) &=
   \int \tfrac{dt}{2\pi} e^{i\omega t} \chi(t) \notag \\
 &= \sum_{a,b} (\rho_a - \rho_b)\, |\Szimp|^2_{ab}\,\delta(\omega - E_{ab})
\text{,}\label{app:motiv:chid:spec}
\end{align}
with $ \chi(t) \equiv \langle [\Sz( t), \Sz] \rangle _{T} \equiv 
\chi^>(t) - \chi^<(t) $, corresponding to the two terms of the 
commutator, respectively. The last line in \Eq{app:motiv:chid:spec} 
provides the Lehmann representation of $\chi''(\omega)$, with $a$ and 
$b$ complete many-body eigenbasis sets, having $\rho_a=\tfrac{1}{Z} 
e^{-\beta E_a}$ and $E_{ab} \equiv E_b-E_a$. Hence with $\chi(\omega) 
= \chi'(\omega) - i\pi\chi''(\omega)$, the static spin susceptibility 
$\chi(T)$ is obtained through Kramers-Kronig relations (Hilbert 
transform), 
\begin{align}
  \chi(T) = \lim_{\omega\to 0}
  \mathrm{P} \int \tfrac{\chi''(\omega')}{\omega-\omega'}\, d\omega'
 = - \mathrm{P} \int \tfrac{\chi''(\omega')}{\omega'}\, d\omega'
\text{,}\label{app:motiv:chid:T}
\end{align}
with $\mathrm{P}$ indicating principal value integral [for finite 
discrete systems, this skips all energetically degenerate terms in 
\Eq{app:motiv:chid:spec} with $E_a=E_b$; the implications of the 
terms $E_a=E_b$ for finite-size systems or for preserved operators 
are discussed in \App{app:finitesize}]. Note that even though $\chi_0 
\equiv \lim_{T\to0}\chi(T)$ describes a low-energy property, through 
\Eq{app:motiv:chid:T}, it requires dynamical information from all 
frequencies. In contrast, the mixed impurity susceptibility in 
\Eq{chi:fs} results in the plain expectation value $\chifs(T) = \beta 
\langle \Sztot \Szimp \rangle _{T}$. At $T=0$, this corresponds to a 
ground-state expectation value. Consequently, this quantity is static 
and does not explore the dynamics of the system, and hence strictly 
focuses on the low-energy sector. For this reason, as pointed out in 
the main text, this quantity exactly reflects, for example, the 
phase-shifts experienced by the electrons of the bath in the 
low-energy fixed point spectrum. 

Nevertheless, this mixed impurity susceptibility is still
insufficient for the evaluation of a proper scale-preserving 
susceptibility. In order to proceed, while still insufficient, it is
instructive to consider the effects of spectral moments (next
section). This will be followed by the actual motivation of the
scale-preserving susceptibility based on the plain non-interacting
resonant level model.

\subsection{Effects of spectral moments}

The Kramers-Kronig or Hilbert transform in \Eq{app:motiv:chid:T}, in 
a sense, corresponds to the spectral moment with $n=-1$ [by using the 
spectral weight $(\omega')^{n}$ within the integral]. This clearly 
weights small frequencies more strongly. Hence this emphasizes the 
low-energy sector while, nevertheless, it weakly reaches out towards 
large energies. This becomes more pronounced still for $n=0$, which 
simply corresponds to the spectral sum rule, 
\begin{align}
 I \equiv \int \chi^>(\omega')\, d\omega'
&= \int (1-f(\omega'))\chi''(\omega')\, d\omega' \notag \\
&= \langle (\Szimp)^2 \rangle_T \lesssim \tfrac{1}{4}
\text{,}\label{app:motiv:chid:srule}
\end{align}
with $f(\omega)$ the Fermi function. For $T=0$, this exactly 
describes the area underneath the spin-spin correlation function 
$\chi''(\omega)$ for positive or, up to a sign, for negative 
frequencies [\cf \Fig{fig:chi:dyn}; the integral over the entire 
$\chid(\omega)$ for all frequencies yields zero by the antisymmetry 
of $\chid(\omega)$]. 

For the SIAM in the local-moment (Kondo) regime, the value of the 
integral in \Eq{app:motiv:chid:srule} at $T=0$ is close to its upper 
bound, $I_0^\mathrm{SIAM}\lesssim 0.25$, with minor variations of 
$\lesssim 10\%$ depending on the specific model parameters. For the 
Kondo model (which represents the large-$U$ limit of the SIAM, \ie 
$U\gg D$), by construction, the sum-rule in \Eq{app:motiv:chid:srule} 
exactly yields the upper bound $I_0^\mathrm{Kondo} = 1/4$. 

At $T=0$, the scaling of the spectral data $\chi''(\omega)$ by 
$\chi_0 = \lim_{T\to0}\chi(T)$ ensures that the height of 
$\chi''(\omega)$ is properly normalized [\eg see \Fig{fig:chi:dyn}, 
all panels]. Since the area underneath $\chi''(\omega)$ is (roughly) 
conserved, scaling of the frequency $\omega$ by $\chi_0^{-1}$ leads 
to approximate scaling (left panels of \Fig{fig:chi:dyn}). 
Specifically, since for the Kondo model, the area is exactly 
preserved (see above), the remaining horizontal variations in 
\Fig{fig:chi:dyn}(c) \emph{must be due to finite bandwidth}. In 
conclusion, the sum-rule in \Eq{app:motiv:chid:srule} is not 
particularly useful for a proper scale-preserving local 
susceptibility. This is not surprising, considering that it 
represents the spectral moment $n=0$, and hence is strongly 
susceptible to effects of finite bandwidth (for the Kondo model this 
means that, while the area in \Eq{app:motiv:chid:srule} is preserved, 
there can be a shift of spectral weight from the band edge to 
low-energy Kondo regime and vice versa, hence spoiling scaling of the 
low-energy Kondo features). Higher spectral moments will make things 
even worse. Hence this route appears ill-suited for the search of a 
scale-preserving local susceptibility at $T=0$. 

\subsection{Motivation through the non-interacting SIAM
\label{app:motiv:nonint}}

The scale-preserving susceptibility proposed in the main text was
also tested successfully for the asymmetric SIAM, as well as in the
limit $U\to0$ at finite $\Gamma$, \ie the plain non-interacting 
resonant level model. Even there, the proposed $\chisc_0$ still
nicely allowed for scaling of low-energy features, such as the
impurity spectral function $A(\omega)\equiv
-\tfrac{1}{\pi}\mathrm{Im} G_d(\omega)$, as long as the low-energy 
scale (here $\Gamma$) is clearly smaller than the bandwidth, \ie
$\Gamma \lesssim 10^{-2}$. The reason for this will be explained in
what follows.
Considering that the general impurity Green's function for an 
interacting system can be written as $G_d(\omega) = [ \omega - 
\varepsilon_d - \Delta(\omega) - \Sigma(\omega)]^{-1}$, with 
$\Sigma(\omega)$ the impurity self-energy, the discussion of the 
effects of finite bandwidth on the hybridization function 
$\Delta(\omega)$ below may serve as a more general motivation, 
indeed, for the definition of a scale preserving susceptibility. In 
particular, as it is demonstrated in the main paper, the result can 
also be nicely applied to interacting systems. 

For the non-interacting case, with $\sigma \in\{\uparrow,
\downarrow\} \equiv \{\pm 1\}$, the spin susceptibility reduces to
the impurity charge-susceptibility for the spinless model. With
$\langle \Szimp \rangle_T=0$, one has
\begin{subequations}
\begin{align}
   \chid(T) &= \tfrac{1}{4}
     \sum_{\sigma,\sigma'} \sigma\sigma' \cdot
     \underset{
        \propto \delta_{\sigma\sigma'} }{\underbrace{
        \langle \hat{n}_\sigma\Vert \hat{n}_{\sigma'} \rangle_0
     }}
   = \tfrac{1}{2}
     \langle \hat{n}_{(\sigma)}\Vert \hat{n}_{(\sigma)} \rangle_0
     \notag\\
  &\equiv -\tfrac{1}{2} \lim_{\omega\to0} \chic(\omega)
\text{,}\label{app:motiv:nonint:suscept}
\end{align}
[regarding the sign in the last line, see \Eq{app:motiv:chid:T}],
with the charge susceptibility given by
\begin{align}
   \chic(\omega) \equiv \mathrm{FT}\bigl(
      -i \vartheta(t)
      \langle [\hat{n}(t), \hat{n}]\rangle_T
   \bigr)
\text{,}\label{app:motiv:nonint:ncorr:0}
\end{align}
\end{subequations}
with $\hat{n} \equiv \hat{d}^\dagger \hat{d}$, and $\mathrm{FT}(\,)$
indicating Fourier transform. In the non-interacting case, this
results in the impurity susceptibility
\begin{align}
   \chid(T)
 = -\tfrac{\partial}{\partial\varepsilon_d}\langle \hat{n} \rangle_T
 = \mathrm{Im} \int \tfrac{d\omega}{2\pi}
   [G_d(\omega)]^2 f(\omega)
\text{,}\label{app:motiv:nonint:ncorr}
\end{align}
with $G_d(\omega)$ the impurity Green's function and $f(\omega)$ the
Fermi function. This results in the correct large temperature limit, 
$\lim_{T\to\infty} T\chi_0(T)=\tfrac{1}{8}$ for arbitrary
$G_d(\omega)$. The low-temperature limit is model dependent.
Considering the non-interacting case, the impurity Green's function
is given by $G_d(\omega) = [\omega - \varepsilon_d -
\Delta(\omega)]^{-1}$, with $\Delta(\omega^+) \equiv \sum_k
\tfrac{V_k^2}{\omega^+-\varepsilon_k} \equiv E(\omega) - i
\Gamma(\omega)$ the hybridization function. In the wide-band limit
for constant $\Gamma(\omega) =\theta(D-|\omega|)\Gamma$, it follows
that $E(\omega)\to 0$. The effects of finite bandwidth $D$ manifest
themselves at small frequencies $\omega$ through
\begin{subequations}
\begin{align}
    \varepsilon_d \quad\to\quad \varepsilon_d + E(\omega) \simeq
    \tilde{\varepsilon}_d - a \omega
\text{,}\label{app:motiv:nonint:self}
\end{align}
with $\tilde{\varepsilon}_d \equiv \varepsilon_d +E(0)$ and $a\equiv 
- \left. \tfrac{d}{d\omega}E(\omega)\right|_{\omega=0} \sim \Gamma/D 
\ll 1$ some dimensionless small constant (note that for the 
particle-hole symmetric resonant level model with constant $\Gamma$, 
one has $a\ge 0$). This leads to the scaling 
\begin{align}
  \omega \to \tilde{\omega}\equiv
  (1-a)\omega
\text{.}\label{app:nonint:scaling:w}
\end{align}
of the frequency in $G_d(\omega)$ in \Eq{app:motiv:nonint:ncorr} 
(interestingly, this may be interpreted more generally in an 
interacting context as the scaling of frequency by the quasi-particle 
weight $z$\cite{Shastry13}). Therefore far away from the bandwidth, 
$|\omega|\ll D$, the impurity spectral function appears slightly 
\emph{stretched} along the frequency axis while preserving its 
height. Overall, however, the line shape for small frequencies 
\emph{remains unaltered} up to proper scaling factors. 

With respect to frequency, \Eq{app:nonint:scaling:w} suggests the 
\emph{increased} energy scale $\TKsc = \TKinf/(1-a)$ relative to
\TKinf which, to lowest order in $a$, represents the energy scale in
the wide-band limit. Remembering that $\chi_0\propto\TK^{-1}$
represents an inverse energy scale, one obtains
\begin{align}
   \chisc_0(D) = (1-a)\chiinf_0
\text{,}\label{app:nonint:chisc:chiinf}
\end{align}
\end{subequations}
with $\chisc_0(D)$ the scale-preserving local susceptibility at
given finite bandwidth, and $\chiinf_0 \equiv 1/(4\TKinf)$.

On the other hand, at $T=0$, the Fermi function in 
\Eq{app:motiv:nonint:ncorr} is unaffected by the scaling $\omega \to
\tilde{\omega}$, such that the overall integral in
\Eq{app:motiv:nonint:ncorr} may be rewritten in terms of
$\tilde{\omega}$, resulting in
\begin{align}
    \chid_0(D) \simeq \tfrac{1}{1-a} \chiinf_0
    \overset{\mathrm{\mbox{(\ref{app:nonint:chisc:chiinf})}}}{=}
    \bigl( \tfrac{1}{1-a} \bigr)^2 \chisc_0(D)
\text{.}\label{app:motiv:nonint:scaling1}
\end{align}
With $a>0$, this shows that $\chid_0(D)$ overestimates the
scale-preserving susceptibility $\chisc(D)$ for given finite
bandwidth $D$.

The mixed susceptibility now allows to determine and subsequently
eliminate the scale factors $(1-a)$. With
\begin{subequations}
\begin{align}
    \chifs(T) &= \int_0^\beta d\tau\, \langle \Szimp(\tau) \Sztot \rangle
   = \beta \langle \Szimp \Sztot \rangle \notag\\
  &= \tfrac{\beta}{2} \bigl(
       \langle \hat{n} \hat{N} \rangle -
       \langle \hat{n} \rangle
       \langle \hat{N} \rangle
     \bigr)
\text{,}\label{app:motiv:nonint:chifs}
\end{align}
the last line again already refers to a spinless model, with $\hat{n} 
\equiv \hat{d}^\dagger \hat{d}$ the number of particles at the 
impurity and $\hat{N}$ the total number of particles in the system. 
In the non-interacting case with $A(\omega) \equiv 
-\tfrac{1}{\pi}\mathrm{Im}G_d(\omega)$ the impurity spectral 
function, this becomes 
\begin{align}
    \chifs(T) &= \tfrac{1}{2} \int d\omega A(\omega) (-f'(\omega))
\text{.}\label{app:motiv:nonint:chifs2}
\end{align}
\end{subequations}
In the limit $T\to0$, this yields $\chifs_0=A(0)/2$. While 
$A(\omega)$ depends on the rescaled frequency $\omega \to 
(1-a)\omega$, as discussed above, this is irrelevant here since 
$A(\omega)$ is evaluated at $\omega=0$. In the wide-band limit of a 
featureless bath, \ie constant hybridization $\Gamma$, 
\Eq{app:motiv:nonint:ncorr} exactly agrees with 
\Eq{app:motiv:nonint:chifs2}. Together with the fact that $\chifs_0$ 
does not explicitly depend neither on the bandwidth nor dynamically 
on finite frequency, this allows to identify $\chifs_0 = \chiinf_0$ 
even at \emph{finite} $D$. 

Using \Eq{app:motiv:nonint:scaling1}, the effects of finite
bandwidth on $\chisc_0(D)$ to lowest-order in $a$ are thus
summarized by
\begin{align}
   \chisc_0(D) = (1-a)^2 \chid_0(D)
   \simeq (1-2a)\chid_0(D)
\text{.}\label{app:motiv:nonint:scaling2}
\end{align}
The first reduction of $\chid_0(D)$ by the factor $(1-a)$ leads to 
$\chifs_0$. Another reduction by the same factor leads to the desired 
$\chisc_0(D)$. With $a\ll1$, this implies that the difference between 
$\chid_0(D)$ and $\chifs_0$, as well as the difference between 
$\chifs_0$ and $\chisc_0(D)$ are the same to lowest order in $a$, and 
are given by the first equality in \Eq{app:motiv:nonint:scaling1}, 
$a\chid_0(D) \simeq \chid_0(D) - \chifs_0$. Together with the last 
term in \Eq{app:motiv:nonint:scaling2} then, one obtains the final 
expression for the scale-preserving local susceptibility, 
\begin{align}
   \chisc_0(D) 
&= 2\chifs_0 - \chid_0(D)
\text{,}\label{app:motiv:sc:final}
\end{align}
in agreement with \Eq{def:chi0:b} in the main paper.

\section{Impurity susceptibility and finite size effects
\label{app:finitesize}}

Consider the Lehmann representation of the generic impurity
susceptibility given by the last term in \Eq{XY:suscept:1}, 
\begin{subequations}
\label{XY:suscept:ab}
\begin{align}
& \SPTh{X}{Y}
= \sum_{a,b} \tfrac{e^{-\beta E_{a}}}{Z} (\delta X)_{ab}(\delta Y)_{ba}
  \tfrac{1-e^{-\beta E_{ab}^{+}}}{E_{ab}^{+}}
\label{XY:suscept:ab:1} \\
& =
  \underset{
    =\SPTh[R]{X}{Y}
  }{ \underbrace{\sum_{a\neq b}
     \tfrac{e^{-\beta E_{a}}-e^{-\beta E_{b}}}{Z}
     \tfrac{X_{ab}Y_{ba}}{E_{ab}^{+}}}
  }
+ \underset{
     \equiv \SPTh[\delta]{X}{Y}
  }{ \underbrace{\beta \sum_{a}
     \tfrac{e^{-\beta E_{a}}}{Z}( \delta X) _{aa}( \delta Y) _{aa}}}
\text{.}\label{XY:suscept:ab:2}
\end{align}
\end{subequations}
Here $a$ and $b$ represent complete many-body eigenbasis sets, i.e. 
$\hat{H} \vert a\rangle =E_{a}\vert a\rangle $ with $ E_{ab}\equiv 
E_{b}-E_{a}$, and the Boltzmann distribution $\rho _{a}=e^{-\beta 
E_{a}}/Z$ (note that $(\delta X)_{aa}=X_{aa}-\langle \hat{X} \rangle 
_{T}\neq 0$ in general). In the first line the positive 
infinitesimal, $E_{ab}^{+} \equiv E_{ab}+i0^{+}$, was added for 
convenience to correctly deal with the case $E_{a}=E_{b}$ (the sign 
of the infinitesimal imaginary part is initially actually irrelevant 
here). By splitting off the terms $a=b$ of the sum in 
\Eq{XY:suscept:ab:1} into the correction $\SPTh[\delta]{X}{Y}$, the 
first term in \Eq{XY:suscept:ab:2} then translates into the Kubo 
formula for linear response $\SPTh[R]{X}{Y}$ based on the retarded 
response function. By the way the specific infinitesimals are chosen, 
actually all degenerate terms $E_{a}=E_{b}$ drop out of the first 
term (principal value integral in the continuum's limit), which 
therefore ignores accidental degeneracies, \ie degeneracies beyond 
strict internal multiplet degeneracies due to symmetry which are 
included with the second term. As a consequence, the sum in the first 
term can be relaxed back to all $a,b$ including $a=b$. Furthermore, 
the correction $\SPTh[\delta]{X}{Y}$ in \Eq{XY:suscept:ab:2} is 
relevant only if the spin states of the states $a$ are sufficiently 
long-lived. In the extreme case $\hat{X} = \hat{Y} = 
\hat{S}_{z}^{\mathrm{tot}}$, the first term $\SPTh[R]{X}{Y}$ in 
\Eq{XY:suscept:ab} is strictly zero, and therefore the entire 
susceptibility is carried by the second term. In contrast, for the 
case that the Hamiltonian does not commute with $\hat{X}$ say, in the 
thermodynamic limit one expects that $X_{aa}\rightarrow 0$ and the 
second term in \Eq{XY:suscept:ab} vanishes. In this case linear 
response is safe using either Kubo formula or the imaginary-time 
Matsubara susceptibility. However, in the presence of discretized 
finite-size systems, $X_{aa}\neq 0$ can become a significant 
contribution nevertheless! In this case, both contributions in 
\Eq{XY:suscept:ab} must be included. 

\subsection{Limit of large temperature for finite system}

For a finite system in the limit $\beta \vert E_{ab}\vert \ll 1$,
\Eq{XY:suscept:ab:1} becomes
\begin{align}
&  \lim_{T\rightarrow \infty }\SPTh{X}{Y} \simeq
   \sum_{a,b} \tfrac{e^{-\beta E_{a}}}{Z} (\delta X)_{ab}(\delta Y)_{ba}
   \underset{=\beta}{
      \underbrace{\tfrac{1-( 1-\beta E_{ab}^{+}) }{E_{ab}^{+}}}
   }
\notag \\
&= \beta \lim_{T\rightarrow \infty }
   \langle \delta \hat{X}\cdot \delta \hat{Y} \rangle _{T}
 = \beta \lim_{T\rightarrow \infty } [ \langle \hat{X}\hat{Y} \rangle_T
 - \langle \hat{X}\rangle _{T} \langle \hat{Y} \rangle _{T} ]
\text{,}\label{susc:large:T}
\end{align}
which is equivalent to the situation where either operator $\hat{X}$
or $ \hat{Y}$ actually commutes with the Hamiltonian! This again
serves to emphasize the importance of both terms in the evaluation of
the impurity susceptibility in \Eq{XY:suscept:ab} in any numerical
setting for a finite system, even if both, $\hat{X}$ and $\hat{Y}$,
do not commute with the Hamiltonian. While in the case of small $T$
the last term in \Eq{XY:suscept:ab:2} may be negligible, it gains 
relative importance with increasing temperature, to the extent, that
for a finite system with $T\rightarrow \infty $ comparable weight is
carried by both terms in \Eq{XY:suscept:ab:2} [note that for large
$T$, $\SPTh[R]{X}{Y} \propto 1/T$, while the $1/T$ behavior of the
correction $\SPTh[\delta]{X}{Y}$ is caused by the leading $\beta$;
\cf explicit NRG analysis in \Fig{fig:gg:parts} below].

\subsection{Impurity susceptibility at large temperatures}

In the limit $T\rightarrow \infty $, the thermal density matrix is 
fully mixed and hence independent of the eigenbasis of the actual 
Hamiltonian. The thermal average therefore can be reduced to the 
thermal average within the impurity space alone. Therefore with 
$\hat{S}_{z}^{\mathrm{tot}}\equiv \sum_{n}\hat{S} _{z}^{( n) }$ 
summed over all (Wilson) sites $n$ including the impurity, having 
$\langle \Szimp\rangle_T = 0$, \Eqr{def:chi0:1}{chi:fs} reduce to the 
same asymptotic form 

\begin{eqnarray}
   (T\chi)_{\infty }&\equiv& \lim_{T\rightarrow \infty } T \chisc(T)
   \simeq \lim_{T\rightarrow \infty } \SPTn{\Szimp}{\Szimp} \notag \\
&=&
   \tfrac{1}{d_i} \sum_{\sigma_i } \left( S_{z,\sigma_i }^\imp \right)^{2}
\text{,}\label{app:chi:largeT}
\end{eqnarray}
where the impurity is described by the state space $\sigma_i $ of 
dimension $d_i$ that also diagonalizes $\hat{S}_z^\mathrm{d}$. For a 
Kondo impurity, or also for an Anderson impurity in the case $\TK\ll 
D\ll T\ll U$, this implies $\chi _{\infty}= \tfrac{1}{4T}$ [this also 
may be taken as a motivation for the definition of the Kondo 
temperature $\TK=\tfrac{1}{4\chi _{0}}$ in \Eq{def:TK:chi0} in the 
opposite limit of $T\rightarrow 0$; more generally still, for an 
impurity of spin $S$ one obtains $(T\chi)_{\infty} = 
\frac{S(S+1)}{3}$]. On the other hand, for an Anderson impurity with 
$T\gg U$, one obtains $\chi _{\infty}=\tfrac{1}{8T} $ due to the 
enlarged accessible local state space \cite{Merker12} [see also 
\Figsp{fig:chi:gg}{a-b}].

\subsection{Implications for the NRG}

Above considerations are clearly relevant for numerical simulations 
such as the NRG. There the effective length of the Wilson chain 
becomes ever shorter for calculations with increasing temperature 
(automatically so in case of fdm-NRG).\cite{Wb07,Wb12_FDM} In case of 
NRG, the interplay between finite-size effects and large temperatures 
can therefore be considered enhanced. 

\begin{figure*}[tbh!]
\begin{center}
\includegraphics[width=1\linewidth]{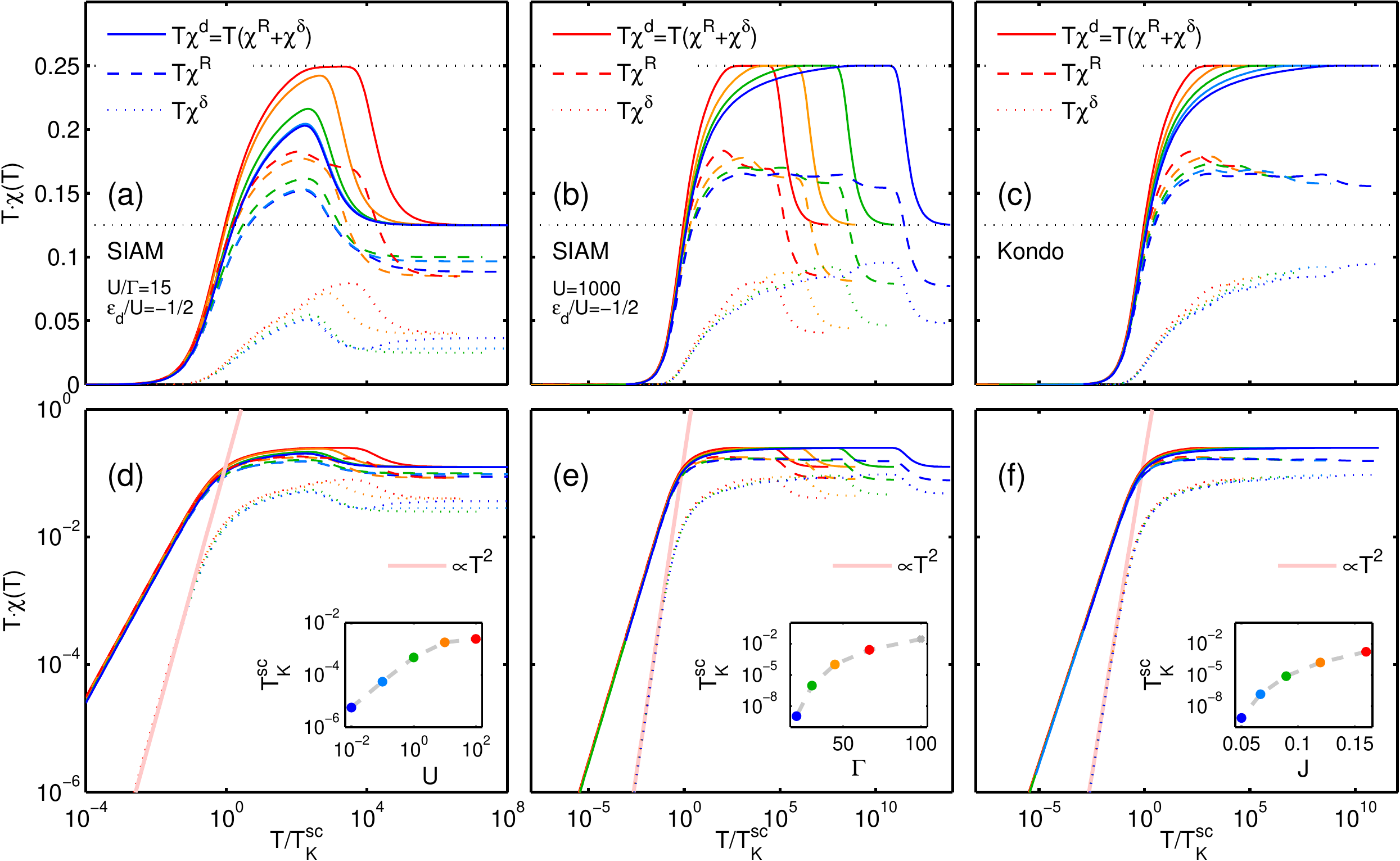}
\end{center}
\caption{ (Color online)
   Contributions to the impurity susceptibility \chid
   as in \Eq{XY:suscept:ab} for the data in \Fig{fig:chi:gg}
   in the main paper
   [panels (a-c) have exactly the same parameter setting
   as \Fig{fig:chi:gg}(a-c)]. The lower panels replicate
   the same data as in the upper panel, yet switching
   to a log-scale also on the vertical axis. The thick light
   solid line corresponds to a plain power-law fit, suggesting
   that the correction $T\chi^{\delta}$ decays like $1/T^2$,
   hence becomes irrelevant in the limit $T\to 0$.
   The insets in the lower panels have been replicated
   from \Fig{fig:chi:gg} to indicate the parameter setting.
}
\label{fig:gg:parts}
\end{figure*}

The two contributions to the static susceptibility in 
\Eq{XY:suscept:ab} are analyzed in detail in \Fig{fig:gg:parts} for 
the data in \Fig{fig:chi:gg} of the main paper. From the log-log 
plots in the lower panels it is clearly seen that $T\chi^{\delta} 
\propto 1/T^2$ for $T\ll \TK$ [in contrast to $T\chi^{R} \propto 
1/T$], and hence becomes negligible in the limit $T\to0$. 
Nevertheless, once $T$ increases and becomes comparable to \TK, the 
correction $T\chi^{R}(T)$ becomes sizable. While the two 
contributions to the static susceptibility in \Eq{XY:suscept:ab} show 
rather irregular behavior individually, as seen in 
\Fig{fig:gg:parts}, their sum yields a smooth physically meaningful 
curve. 

In practice, when computing the first term in \Eq{XY:suscept:ab:2} as 
standard susceptibility within linear response (Kubo formula), the 
second term shows up in a disguised manner as $\delta(0)$ 
contribution with opposite sign for $\omega=0^\pm$. This may be 
collected in the smallest frequency bin for positive and negative 
frequencies, respectively, when collecting the discrete data. While 
these $\delta(0)$ contributions drop out of the principal value 
summation in the Kramers-Kronig transformation, nevertheless, it it 
represents, and thus can be simply used to subsequently evaluate the 
correction given by the last term in \Eq{XY:suscept:ab:2}. 

\section{Calculation of the mixed susceptibility
$\protect\chifs(T)$ within fdm-NRG \label{app:chifs:NRG}}

Given that the total spin operator $\hat{S}_{z}^{\mathrm{tot}}$ 
commutes with the Hamiltonian, the mixed susceptibility $\chifs(T) 
\equiv \langle \Szimp \Vert \hat{S}_{z}^{\mathrm{tot}}\rangle _{T}$ 
in \Eq{chi:fs} can be evaluated in a simple and cheap manner, as it 
reduces to the plain set of expectation values, $T\chifs(T) = \langle 
\hat{S}_{z}^{\mathrm{tot}} \Szimp \rangle _{T}-\langle 
\hat{S}_{z}^{\mathrm{tot}}\rangle _{T}\langle \Szimp \rangle _{T}$. 
This includes one local operator $\Szimp$ and one global operator, 
the total spin operator $\hat{S}_{z}^{\mathrm{tot}}\equiv \sum_{n} 
\hat{S}_{z}^{( n) }$ which is given by the sum of local spins 
$\hat{S}_{z}^{( n) }$ associated with site $n$ along the Wilson chain 
including the impurity, say, at $n=-1$. Being interested in the 
magnetic susceptibility at zero magnetic field, it follows $\langle 
\hat{S}_{z}^{\mathrm{tot}} \rangle _{T}=\langle \Szimp \rangle 
_{T}=0$. The remaining quantity then, 
\begin{equation}
   T\cdot\chifs( T) = \langle \hat{S}_{z}^{\mathrm{tot}}\Szimp \rangle _{T}
 = \mathrm{\mathrm{tr}}[
     \hat{\rho}( T) \cdot
     \hat{S}_{z}^{\mathrm{tot}}\Szimp
   ]
\text{,}  \label{app:chi:mixed}
\end{equation}%
is a simple intrinsic quantity that is solely related to the 
impurity. In given case only a single sum over a complete many-body
eigenbasis $a$ suffices, with the Lehmann representation of
\Eq{app:chi:mixed} given by
\begin{equation}
   T\cdot\chifs( T) = \sum_{a}\tfrac{e^{-\beta E_{a}}}{Z}
   S_{z,a}^{\mathrm{tot}}\left( \Szimp \right) _{aa}
\text{,} \label{app:chi:mixed:LM}
\end{equation}%
where $\hat{H}\vert a\rangle \equiv E_{a}\vert a\rangle $. By
construction, the full thermal density matrix as well as the total
spin operator $S_{z}^{\mathrm{tot}}$ are strictly diagonal, with the
matrix elements given by $\left[ S_{z}^{\mathrm{tot}}\right]
_{aa^{\prime }}=\delta _{aa^{\prime }}S_{z,a}^{\mathrm{tot}}$ and
$\left[ \hat{\rho}( T) \right] _{aa^{\prime }}=\delta _{aa^{\prime
}}e^{-\beta E_{a}}/Z$, respectively, with $Z( T) \equiv 
\sum_{a}e^{-\beta E_{a}}$ the grand-canonical partition function.

In what follows, the complete basis set $a$ is given by the
iteratively discarded state spaces generated by the NRG,
\cite{Anders05} i.e. $\vert a\rangle \rightarrow \vert se\rangle
_{n}^{D}\equiv \vert s\rangle _{n}^{D}\otimes \vert e\rangle _{n}$
with $s_{n}\in D$ a discarded state at iteration $n$ and $e_{n}$ the
environment \wrt iteration $n$, i.e. the full state space for the 
remainder of the Wilson chain $n\,<n^{\prime }\leq N$ with $N$ the
final length of the Wilson chain considered. The resulting full
thermal density matrix (fdm) is given by \cite{Wb07,Wb12_FDM}
\begin{equation}
   \hat{\rho}( T) =
   \sum_{n}w_{n}(T) \hat{\rho}_{n}^{D}( T)
\text{,}\label{def:FDM}
\end{equation}%
where $w_{n}( T) $ is a well-defined temperature-dependent weight 
distribution along the Wilson chain that is peaked near the energy 
scale of temperature. The operators $\hat{\rho}_{n}^{D}$ are 
normalized thermal density matrices within the discarded state space 
of iteration $n$ (the sum over the environment of the remaining 
iterations, resulting in the degeneracy factor $d^{N-n}$ with $d$ the 
dimension of the local state space of a single Wilson site, has been 
already properly included in the weight distribution $w_{n}$). 
\cite{Wb07,Wb12_FDM} With the full thermal density matrix a scalar 
operator, all entries in \Eq{def:FDM} are block-diagonal. In 
particular, being initialized within the discarded (eigen-) state 
space at iteration $n$ itself, all $\hat{\rho}_{n}^{D}$ are strictly 
diagonal. 

Now, assuming that also $\hat{S}_{z}^{\mathrm{tot}}$ commutes with 
the Hamiltonian, it is also block diagonal. Using the complete basis 
set $\vert se\rangle _{n}^{D}\equiv \vert s\rangle _{n}^{D}\otimes 
\vert e\rangle _{n}$, in the expectation value in 
\Eq{app:chi:mixed:LM} for the mixed susceptibility, the environment 
is traced over. Specifically with 
\begin{equation*}
   \hat{S}_{z}^{\mathrm{tot}}\equiv \sum_{n}\hat{S}_{z}^{( n) }=
   \underset{
       \equiv \hat{S}_{z}^{n,\mathrm{tot}}
   }{\underbrace{
       \sum_{n^{\prime }\leq n}\hat{S}_{z}^{( n^{\prime }) }}
   }
 + \underset{
      \equiv \hat{S}_{z}^{e,\mathrm{tot}}
   }{\underbrace{
      \sum_{n^{\prime }>n}^{N}\hat{S}_{z}^{( n^{\prime }) }}
   }
\text{,}
\end{equation*}
the total spin of the entire Wilson chain splits into two parts \wrt 
a given iteration $n$, the total spin up to and including site $n$, 
and the total spin for the remainder of the chain. The corresponding 
matrix elements are given by (note that the degeneracy factor 
$d^{N-n}$ has been already included with the weight distribution 
$w_{n}$ and is thus compensated in the following expression), 
\begin{align*}
 & \tfrac{1}{d^{N-n}}
   \sum_{e_{n}}\langle se\vert \hat{S}_{z}^{\mathrm{tot}}
   \vert s^{\prime }e\rangle _{n} = \\
 &=\delta _{ss^{\prime }}S_{z,s}^{n,\mathrm{tot}}
  +\delta _{ss^{\prime}}\sum_{n^{\prime }>n}
   \underset{= \langle \hat{S}_{z}^{( n')}\rangle_{\infty} = 0}{\underbrace{
      \tfrac{1}{d} \sum_{\sigma _{n^{\prime }}}
      \langle \sigma _{n^{\prime }}
      \vert \hat{S}_{z}^{( n^{\prime }) }
      \vert \sigma _{n^{\prime }}\rangle
   }}
\text{,}
\end{align*}
where $\sigma _{n^{\prime }}$ spans the $d$-dimensional local Hilbert 
space of Wilson site $n^{\prime }$. The last term represents the 
fully mixed average of the local spin for a given site $n'$, \ie 
corresponding to an effective $T=\infty$, and thus vanishes 
identically by symmetry. Overall, this implies that at iteration $n$, 
only the total spin $\hat{S}_{z}^{n,\mathrm{tot}}$ up to and 
including site $n$ needs to be considered. Therefore the mixed 
susceptibility in \Eq{app:chi:mixed} can be evaluated in the NRG 
context as follows, 
\begin{equation}
   T\cdot\chifs(T) = \sum_{n}w_{n}( T)
   \underset{
      =\sum\limits_{s\in D_{n}}\rho _{n,s}( T)
       S_{z,s}^{n,\mathrm{tot}}\left( S_{z}^\imp \right)_{ss}
   }{\underbrace{
       \mathrm{\mathrm{tr}}\left[ \rho_{n}^{D}( T)
       S_{z}^{n,\mathrm{tot}}S_{z}^\imp \right]
   }}
\text{,}\label{app:chi:mixed:NRG}
\end{equation}
where the trace runs over the discarded state space of iteration $n$ 
as indicated. Here the notation of the operators without hats 
indicates that they already correspond to the matrix representations 
in the basis $s\in D_{n}$, \ie the discarded states at iteration $n$. 
The computationally most expensive part for the result 
\Eq{app:chi:mixed:NRG} is the evaluation of the matrix elements of 
$\Szimp$ in the discarded state space of iteration $n$. From these, 
however, only the diagonals are required. Once computed, the 
calculation of $\chifs( T) $ becomes extremely fast for an arbitrary 
set of temperatures. It is important, though, that for the physically 
correct impurity susceptibility thermal averaging at $T=0^+$ is 
required. Hence the Wilson chain has to be chosen long enough such 
that the weight distribution $w_n(T)$ clearly fits within the Wilson 
chain, \ie $w_N(T) \lesssim 10^{-2}$, with $N$ the length of the 
Wilson chain considered (in practice, $T\gg \omega_N$). 

\subsection{Evaluation in the presence of
non-abelian symmetries}

In the above discussion, the external magnetic field was applied in 
the $z$ -direction. However, if the magnetic susceptibility at $B=0$ 
is computed, the Hamiltonian typically possess $SU(2)$ spin symmetry. 
This can be taken advantage of when evaluating the mixed 
susceptibility above as follows. Clearly, the evaluation of the mixed 
susceptibility \Eq{app:chi:mixed} can be symmetrized \wrt $x$-, $y$-, 
and $z$-components, \cite{Wb12_SUN} 
\begin{equation*}
   T\chifs( T) =\langle \hat{S}_{z}^{\mathrm{tot}}\Szimp \rangle_{T}
 = \tfrac{1}{3} \langle \hat{S}^{\mathrm{tot}}\cdot \Szimp \rangle_{T}
\text{,}
\end{equation*}
where $\hat{S}\equiv \lbrack \tfrac{-1}{\sqrt{2}} \hat{S}_{+},
\hat{S}_{z}, \tfrac{+1}{\sqrt{2}} \hat{S}_{-}$ $]^{T}\equiv \{
\hat{S}_{\mu } \} $ with $\mu \in \{ +1,0,-1\} $ represents the
irreducible three-dimensional spinor for the spin operator which
transforms according to a spin $J=1$ multiplet. Now every component
in the spinor $\hat{S}^{\mathrm{tot}}$ commutes with the Hamiltonian
such that $\hat{S}_{\pm }^{\mathrm{tot}}$ only raises or lowers the
state index within the \textit{same} multiplet, but never leaves a
given multiplet. As a consequence, $\hat{S}^{\mathrm{tot}}$ is still 
a strictly diagonal operator in multiplet space, while the
non-diagonal matrix elements within the same multiplet factorize as
Clebsch-Gordan coefficients (\cf Wigner Eckart theorem). To be
specific, in the presence of symmetries, the state space at each
iteration $n$ is organized using the composite index labels
\cite{Wb12_SUN} $\vert s\rangle _{n}\rightarrow |Js;M\rangle _{n}$
where $s_{(J)}$ now labels a specific multiplet within symmetry
sector $J$, and $M_{(J)}$ represents the $S_{z}$ labels, i.e.
sequences the internal state space of multiplet $J$. With this, the
matrix elements of the total spin operators are given by
\begin{align*}
 & \langle J^{\prime }n^{\prime };M^{\prime }\vert
   \hat{S}_{\mu }^{n,\mathrm{tot}}\vert Jn;M\rangle \\
 &=\delta _{JJ^{\prime }}
   \underset{
      \equiv \Vert S_{J}^{n,\mathrm{tot}}\Vert _{nn^{\prime }}
   }{\underbrace{
      \delta _{nn^{\prime }}\sqrt{J( J+1) }
   }}
   \cdot (JM^{\prime }|1\mu ;JM)
\end{align*}
The prefactor in the reduced matrix elements $\Vert
S_{J}^{n,\mathrm{tot}}\Vert $ for symmetry sector $J$ guarantees
that one obtains the familiar Casimir operator,
\begin{equation}
   \langle Jn;M^{\prime }\vert (\hat{S}^{n,\mathrm{tot}})^{\dagger}
   \cdot \hat{S}^{n,\mathrm{tot}}\vert Jn;M\rangle
 = J(J+1) \delta _{MM^{\prime }}
\text{.}\label{app:Stot2}
\end{equation}
Consequently, in the presence of SU(2) spin symmetry, within the NRG 
the mixed susceptibility in \Eq{app:chi:mixed:NRG} can be rewritten 
as follows, 
\begin{equation}
   T\chifs( T) = \tfrac{1}{3} \sum_{n}
   w_{n}\,\mathrm{\mathrm{\mathrm{\mathrm{tr}}}}
   \left[
      \rho_{n}^{D}~\left( S^\imp \cdot S^{n,\mathrm{tot}}\right)
   \right]
\text{.}\label{app:chi:mixed:NRG2}
\end{equation}
The apparent overhead in terms of the extra summation over the $\mu $ 
components of the spinors in $S^\imp \cdot S^{n,\mathrm{tot}}$ is 
completely negligible when compared to the gain by the reduced 
dimensionality on the reduced matrix element, i.e. the multiplet 
level. First of all, it only affects Clebsch-Gordan coefficient 
spaces. Moreover, by inspecting the block-diagonal structure of 
\Eq{app:chi:mixed:NRG2}, for the specific contribution of any 
symmetry sector within the trace \textit{exactly the same} 
Clebsch-Gordan coefficient space appears twice, in both $S_{\mu 
}^{n,\mathrm{tot}}$ as well as $S_{\mu }^\imp$. Hence, by performing 
the trace for the Clebsch Gordan coefficient space similar to 
\Eq{app:Stot2}, this only adds a factor $( 2J+1)$, \ie the 
$3j$-symbol, which is simply equal to the dimensionality of multiplet 
$J$. Hence the explicit contraction of the Clebsch-Gordan 
coefficients can be fully circumvented. In summary, the effect of 
non-abelian symmetries on the evaluation of the mixed susceptibility 
in \Eq{app:chi:mixed:NRG2} is that (i) $S^\imp$ can be reduced to its 
block-diagonal components due to the block-diagonal structure of all 
the remaining participants. (ii) The traced-over Clebsch-Gordan 
spaces together with the definition of $S^{n,\mathrm{tot}}$ results 
in the combined factor $\tfrac{1}{3} \sqrt{J( J+1) }( 2J+1) $ for 
symmetry sector $J$ that can be directly multiplied onto the reduced 
matrix elements of $S^\imp$. Finally, with the Clebsch-Gordan 
coefficients taken care of, (iii) the remaining trace is carried out 
over the reduced multiplet space only.

\subsection{Evaluation of the approximate impurity specific heat
${\protect{\langle (\Himp+\tfrac{1}{2}\Hcpl)\Htot \rangle_T}}$
within fdm-NRG \label{app:specific:heat}}

The impurity specific heat has a similar mathematical structure when 
compared to the general discussion of susceptibility above. However, 
since it would be a susceptibility that refers to the temperature 
itself as the variable physical parameter, in the presence of thermal 
averages, these similarities necessarily remain vague and the 
impurity specific heat is special. Nevertheless, as it turns out, 
\cite{Merker12cv} the impurity specific heat can also be computed 
through the following \emph{local} approximation, 
\begin{eqnarray}
   c_V(T) \simeq
   \tfrac{\partial }{\partial \Ttot} \langle \Hipc \rangle_T
 = \tfrac{\partial }{\partial \Tipc} \langle \Htot \rangle_T
\text{,}\label{app:cv:imp}
\end{eqnarray}
where $\Hipc \equiv \Himp+\tfrac{1}{2}\Hcpl$, with \Himp and \Hcpl 
the impurity Hamiltonian and its coupling to the bath, respectively 
[\eg see \Eq{eq:SIAM}; here \emph{ipc} stands for impurity plus part 
of the coupling to the bath]. The first expression, $\tfrac{\partial 
}{\partial T_{(\mathrm{tot})}} \langle \Hipc \rangle$, has the 
intuitive physical interpretation that it represents the change in 
energy at the impurity due to a change in the overall total 
temperature, where the contribution of the hybridization is shared 
\emph{in equal parts} with the bath \cite{Merker12cv}. 
Mathematically, this is equivalent to the second expression in 
\Eq{app:cv:imp}, $\tfrac{\partial }{\partial \Tipc}\langle \Htot 
\rangle$, which represents the change in total energy due to a change 
in local temperature, \ie with $\beta\equiv 1/\Ttot$ and $\Htotb 
\equiv \Hipc+\Hbpc$ (where \emph{bpc} stands for bath plus remaining 
contribution from the coupling to the impurity), 
\begin{align}
   e^{-\beta\hat{H}} &\equiv \exp\left(-\tfrac{1}{\Ttot}(\Hipc+\Hbpc)\right)
  \notag \\ & \to
   \exp\left(-\tfrac{1}{\Tipc}\Hipc -\tfrac{1}{\Tbpc}\Hbpc\right)
\text{,}\label{ap:interp2}
\end{align}
evaluated at $\Tipc=\Tbpc=\Ttot$ \emph{after} taking the derivative
for $c_V(T)$, as indicated by the trailing subscript $T$ in the last
term of \Eq{app:cv:imp}.

While in [\onlinecite{Merker12cv}] the derivative in \Eq{app:cv:imp} 
was computed numerically by first computing the expectation values
$\langle \Hipc \rangle_T$, the derivative in \Eq{app:cv:imp} can be
easily expressed analytically,
\begin{eqnarray}
   c_{V}(T) &=& \beta ^{2}\Bigl[
      \langle \Hipc \Htot \rangle_T
    - \langle \Hipc \rangle_T \langle \Htot\rangle_T
   \Bigr]
\text{,}\label{app:cvT:2}
\end{eqnarray}
which still can be directly evaluated numerically within the NRG
using complete basis sets. \cite{Anders05, Wb07,Wb12_FDM} The term
$\langle \Hipc \rangle_T$ corresponds to a simple thermal average of
a local quantity. \cite{Wb12_FDM} The total energy, on the other
hand, is given by
\begin{subequations}
\begin{align}
   \langle \Htot \rangle _{T} &= \sum_{n,s\in D}
   \hspace{-0.20in} \underset{=w_{n}(T)
   \tfrac{e^{-\beta E_{s}^{n}}}{Z_{n}}\equiv
   w_{n}\rho _{s}^{n}}{\underbrace{\sum_{e}
   \tfrac{e^{-\beta E_{s}^{n}}}{Z}}} \hspace{-0.20in}
   ( \omega_{n} \tilde{E}_{s}^{n}+\delta _{n} )
\label{app:cvT:2a}
\end{align}
with the eigenenergies $E_{s}^{n} \equiv \omega_{n} \tilde{E}_{s}^{n} 
+ \delta _{n}$ (as is customary, the NRG eigenenergies 
$\tilde{E}_{s}^{n}$ are given in rescaled units, with $\omega_{n}$ 
the energy scale at iteration $n$ and $\delta _{n}$ here the 
\emph{cumulative} subtracted energy offset \wrt the ground state at 
iteration $n$). While a global energy reference drops out of the 
entire definition of the impurity specific heat \Eq{app:cvT:2}, of 
course, the individual energy references $\delta_n$ for Wilson shell 
$n$ \emph{do not} cancel and hence must be properly included. 
Therefore $E_{s}^{n} \equiv \omega_{n} \tilde{E}_{s}^{n}+\delta _{n}$ 
represent the eigenenergies in non-rescaled physical units with 
respect to a single common energy reference, \eg the ground state 
energy of the entire Wilson chain. In this case, the offsets 
$\delta_n$, when computed starting from the low-energy side (\ie 
large $n$) scale like $\delta_n \propto \omega_n$. In 
\Eq{app:cvT:2a}, finally, again a single sum over the complete 
discarded (D) basis set $(s,e,n)^\D$ suffices, since, obviously, 
\Htot commutes with itself, \ie with the Hamiltonian used in the 
evaluation of the overall thermodynamic average. 
With the remaining term in \Eq{app:cvT:2} given by,
\begin{align}
   \langle \Hipc \Htot \rangle _{T} &=
   \sum_{n,s\in D} w_{n}\rho_{s}^{n}
   \left( \omega _{n}\tilde{E}_{s}^{n}+\delta _{n}\right)
   \left\langle s_{n}\right\vert \Hipc \left\vert
   s_{n}\right\rangle
\text{,}\label{app:cvT:2b}
\end{align}
\end{subequations}
the resulting impurity specific heat can be expressed as follows,
\begin{align}
   c_{V}(T) &=
      \beta ^{2}\sum_{n,s\in D} w_{n}\omega _{n}\rho _{s}^{n}
      \tilde{E}_{s}^{n} \Bigl[
         \left\langle s_{n}\right\vert \Hipc \left\vert
         s_{n}\right\rangle -\langle \Hipc \rangle _{T}
      \Bigr]
   \notag \\
&+ \beta^{2}\sum_{n,s\in D} w_{n}\delta _{n}\rho _{s}^{n}
   \Bigl[
      \left\langle s_{n}\right\vert \Hipc \left\vert s_{n}\right\rangle
    - \langle \Hipc \rangle _{T}
   \Bigr]
\notag \\
 & \equiv \sum_{n} w_{n} \Bigl[
       \tfrac{1}{\omega_n}  
       \tilde{c}^{(D,n)}_{V}(T) + \tfrac{\delta_n}{T^2}
       \bigl( \langle \Hipc \rangle_n^{\D} - \langle \Hipc \rangle_T \bigr)
   \Bigr]
\text{,}\label{app:cvT:n}
\end{align}
where $\tilde{c}^{(D,n)}_{V}(T)$ stands for the specific heat 
computed within the discarded states space of Wilson shell $n$ in 
\emph{rescaled} units, \ie using $\tilde{E}_{s}^{n}$ and $T \to 
\tilde{T}_n\equiv T/\omega_n$. While $\tilde{c}^{(D,n)}_{V}(T)$ is 
clearly independent of the energy references $\delta_n$ for each 
individual Wilson shell $n$, these $\delta_n$ do lead to a 
\emph{finite} contribution through the very last term in 
\Eq{app:cvT:n}. The reason is that, in general, the thermal 
expectation value $\langle \Hipc \rangle_n^{\D}$ in the discarded 
state space of iteration $n$ is unequal to the full thermal average 
$\langle \Hipc \rangle_T$ for the entire system. Only for very late 
Wilson shells in the low energy fixed point, \ie $T\to0$, it follows 
$\left\langle s_{n}\right\vert \Hipc \left\vert s'_{n}\right \rangle 
\simeq \langle \Hipc \rangle _{0}\cdot \delta_{ss'}$. This leads to 
cancellation of the last term, which is required for $\lim_{T\to0} 
c_{V}(T) = 0$.

\section{On the extraction of phase shifts within the NRG}
\label{app:phaseshifts}

The Kondo scale \TKfs derived from the mixed susceptibility [see 
\Eq{chi:fs}] is identical to the Kondo scale \TKphi obtained from the 
phase shifts [see \Eq{def:TKphi:0}], \ie $\TKfs = \TKphi$, as 
discussed with \Eq{def:TKphi} in the main text. 
For a Fermi liquid in the thermodynamic limit, the one-particle level 
spacing can be considered equally spaced around the Fermi energy yet 
different for each electronic flavor such as spin $\sigma$, 
\begin{equation}
   \tilde{\varepsilon}_{k\sigma}
 = \epsilon_{1\sigma} + k \cdot \epsilon_{2\sigma}
\text{,}\label{eq:phaseshiftFL:0}
\end{equation}
with $k \in \{\ldots,-2,-1,0,1,2,\ldots\}$ and $\epsilon_{1\sigma} 
\in [0,\epsilon_{2\sigma}[$, given that $\epsilon_{1\sigma}$ is 
essentially defined up to modulo $\epsilon_{2\sigma}$. Here the tilde 
on $\tilde{\varepsilon}_{k\sigma}$ indicates that the original 
decoupled fixed bath modes may already have been shifted by the 
presence of a coupled impurity. If the baths are identical for each 
flavor $\sigma$ including their discretization, $\epsilon_{2\sigma}$ 
is independent of $\sigma$. This is typically the case for NRG where 
$\epsilon_{2\sigma} \propto \omega_N \propto \Lambda^{-N/2}$, with 
$\omega_N$ the energy scale at large but finite length $N$ of the 
Wilson chain. Hence $\epsilon_{1\sigma}/\omega_N$ and 
$\epsilon_{2\sigma}/\omega_N$ are both of order 1. For the ground 
state, all levels with $\tilde{\varepsilon}_{k\sigma} < 0 $ are 
occupied. If $\epsilon_{1\sigma} = 0$, the many-body ground state is 
degenerate. For a Fermi liquid, the phase shift $\varphi_{\sigma}$ 
can be extracted independently for each $\sigma$. In the 
thermodynamic limit, it is given by the ratio 
\begin{equation}
   \frac{\varphi_\sigma}{\pi}
 = \frac{\epsilon_{1\sigma}}{\epsilon_{2\sigma}}
\text{,}\label{eq:phaseshiftFL}
\end{equation}
(this can be simply motivated by using the connection of phase shifts 
to the change in (local) occupation through the Friedel sum rule, 
while taking a proper continuum limit starting from a finite yet 
large system, \ie a discrete model). 

\begin{figure}
\includegraphics[width=1\linewidth]{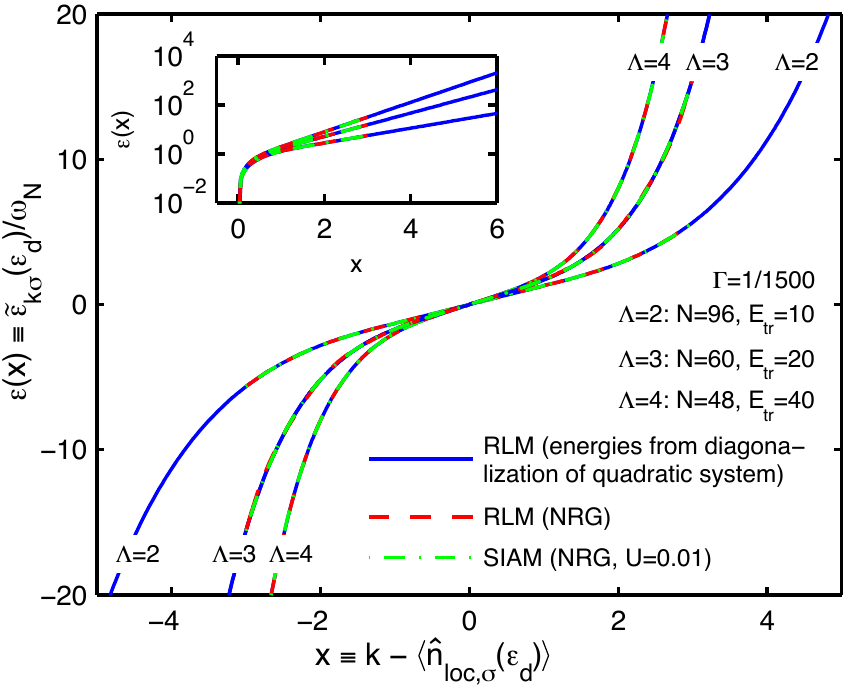}
\caption{(Color online)
  Dependence of single-particle energy level spectra
  $\tilde{\varepsilon}_{k\sigma}(\varepsilon_{d})$ on local occupation
  $\nlocds$ and level index $k$ for the SIAM [NRG (green dot-dashed)]
  as well as the RLM [quadratic solution (blue) and NRG (red
  dashed)] using a long even Wilson chain of length $N$ as specified.
  The local occupation
  $\nlocds$ and thus the phase shift is changed by varying the
  position of the impurity energy level $\varepsilon_{d(,\sigma)}$. While
  this level is swept from $+\infty$ to $-\infty$, $\nlocds$ changes
  smoothly from 0 to 1. Combining all energies in units of the
  energy scale $\omega_n$ \vs $x \equiv k-\nlocds$,
  this results in a single continuous antisymmetric curve
  $\varepsilon(x)$ that is linear for small $|x|$, yet is quickly
  dominated by exponential behavior for larger $|x|\gtrsim 2$ (see
  inset and text).
  The discrete levels $\tilde{\varepsilon}_{k\sigma}(\varepsilon_{d})<0$
  (\ie within the range $x<0$) correspond to single-particle levels
  below the Fermi energy and are thus occupied in the ground state.
  The data for the blue curve was obtained by numerical
  diagonalization of the quadratic Hamiltonian (RLM), hence all
  single-particle energies are easily obtained. In particular, their
  energies are not restricted to the energy range below the truncation
  energy, as is the case for the NRG-method (dashed and dot-dashed
  lines).
}\label{fig:energylevels}
\end{figure}

Within the NRG, the one-particle level position in energy can be 
determined from the many-body eigenspectrum of the energy flow 
diagram, \ie the finite-size fixed-point spectra at $T=0^+$. This 
allows to extract $\varphi_{\sigma}$ through \Eq{eq:phaseshiftFL}. 
Note, however, that due to the intrinsic even-odd alternations with 
the actual shell of the Wilson chain, the resulting phases 
$\varphi_{\sigma}$ differ by the constant offset of $\pi/2$ between 
even and odd shells; nevertheless, since only differences in the 
phases due to the presence of the impurity, \ie phase \emph{shifts}, 
are considered, for an arbitrary but fixed energy shell this offset 
is irrelevant. 
However, \Eq{eq:phaseshiftFL} is based on an equally spaced
one-particle level spectrum around the Fermi energy, which is not
quite the case within NRG at all! Even though NRG does allow to
\emph{directly} access the thermodynamic limit in the numerical 
simulation due to the underlying logarithmic discretization in
$\Lambda$, \cite{Wilson75,Wb11_aoc} for a given length $N$ of the
Wilson chain and a necessarily rather coarse discretization with
$\Lambda \gtrsim 2$, the approximately uniform level spacing around
the Fermi energy quickly transforms into exponentially separated
energy levels further away from the Fermi energy, \cite{Krishna80I}
as shown in \Fig{fig:energylevels}.

\FIG{fig:energylevels} analyzes the single-particle level spectra

for the interacting as well as the non-interacting SIAM [the latter 
also referred to as the resonant level model (RLM)] as defined in 
\Eq{eq:SIAM} for an arbitrary late but fixed even Wilson shell $N$ 
[\ie $H_0$ such as in \Eq{eq:SIAM} plus some larger even number of 
further Wilson sites; for an odd length of the Wilson chain, all 
curves in \Fig{fig:energylevels} would be trivially offset 
horizontally by $1/2$, which can be ignored]. With $\varepsilon_{d} 
\equiv \{\varepsilon_{d\sigma}\}$ the (magnetic field dependent) 
level positions of the impurity, the one-particle level spectrum 
$\tilde{\varepsilon}_{k\sigma} ( \varepsilon_{d})$ of the entire 
system. This shift of the discrete single-particle spectrum for an 
arbitrary but fixed $\varepsilon_{d}$ is directly related to phase 
shifts via Friedel sum-rule. Thus when plotted \vs the 
\emph{continuous} variable $x\equiv k - \nlocds$ having 
$\varepsilon_{d(,\sigma)} \in [ -\infty, \infty ]$ and hence $\nlocds 
\in [0,1]$ with $\nlocds$ the change in local charge at and close to 
the impurity \cite{Muender12} depending on the impurity setting, this 
allows to collect all one-particle level spectra 
$\tilde{\varepsilon}_{k\sigma}( \varepsilon_{d})$ after rescaling by 
the approximate one-particle level spacing $\omega_n$ into a 
\emph{single continuous curve} $\varepsilon(x)$, as demonstrated in 
\Fig{fig:energylevels}. In a sense, with the Wilson chain in mind, 
the presence of the impurity allows to alter the boundary condition 
for the bath electrons, thus resulting in an impurity-dependent phase 
shift, which sets the horizontal offset $\nlocds$ of the discrete 
energy levels in \Fig{fig:energylevels}.

The resulting curve $\varepsilon(x)$, which describes the 
\emph{macroscopic} bath, is universal in the sense that it only 
depends on the bath discretization (\ie $\Lambda$), but is 
independent of the specifics of the \emph{microscopic} impurity as 
long as the low-energy behavior represents an effective Fermi liquid. 
For example, as demonstrated in \Fig{fig:energylevels}, the resulting 
curve $\varepsilon(x)$ is exactly the same independent of whether the 
impurity is interacting (SIAM) or not (RLM, with or without NRG). 
Using the same bath discretization for all flavors $\sigma$, as is 
customary within the NRG, this curve $\varepsilon(x)$ is also 
independent of $\sigma$, as already indicated by its notation.

As a consequence, for a given bath discretization the curve
$\varepsilon(x)$ can simply be computed for the non-interacting case
(spinless RLM) by repeated diagonalization of the underlying
quadratic Hamiltonian while sweeping $\varepsilon_{d} \in [ -\infty, 
\infty ]$ (\eg see solid line in \Fig{fig:energylevels}). With the
NRG bath-discretization being particle-hole symmetric, the resulting
curve $\varepsilon(x)$ is antisymmetric in $x$, \ie $\varepsilon(-x)
= -\varepsilon(x)$.
Then given the reference curve $\varepsilon(x)$ together with the
requirement of its antisymmetry, the single-particle spectrum for
any other impurity setting can be fitted (provided Fermi liquid
behavior), which allows to extract the horizontal offset $\nlocds$
and hence the phase shift $\varphi_\sigma$ independently for each
flavor $\sigma$, \emph{even if} the single-particle spectrum is not
exactly uniformly spaced around the Fermi energy.

The range of linearity of $\varepsilon(x)$ around $x=0$ indicates the 
regime of equally spaced single-particle levels closest to the Fermi 
energy, given an exponentially large but finite system size, as 
represented by the length $N$ of the Wilson chain. For $\Lambda=2$, 
linearity is given to a good approximation (within about 0.8\%) for 
$x\in[-0.5, \, 0.5]$, \ie for the lowest single-particle and 
single-hole excitation in the particle-hole symmetric case, and hence 
justifies using \Eq{eq:phaseshiftFL} [this method was used for 
extracting $\TKphi$ and verifying \Eq{def:TKphi} to within 1\% 
accuracy in the main text]. In contrast, for $\Lambda=4$ the 
linearity of $\varepsilon(x)$ even within this minimal regime is 
already clearly compromised (about 3\%). Here usage of 
\Eq{eq:phaseshiftFL} already leads to clear systematic errors due to 
the strongly increased coarseness of the underlying logarithmic 
discretization, leading to about a 7\% error in \Eq{def:TKphi}. 
Therefore the extraction of phase shifts for larger $\Lambda$ from 
the single-particle spectra requires a more careful analysis such as 
the aforementioned fitting to the curve $\varepsilon(x)$. Given 
logarithmic discretization, it follows that $\varepsilon_{k} \sim 
\mathrm{sgn}(k) \, \omega_N \, \Lambda^{|k|}$ for larger $|k|$ for a 
fixed length $N$ of the Wilson chain. From the semilog-y 
representation in the inset of \Fig{fig:energylevels} it can be seen, 
that for $|x|\gtrsim 2$, $\varepsilon(x)$ is already described by a 
plain exponential behavior to within 0.1\%. Thus rather than fitting 
the data for $|x|\lesssim 1$, alternatively, one may simply 
concentrate on the exponential behavior for larger $|x|$ which, 
however, requires to extract the single particle spectrum at least up 
to the third single particle level. 

\bibliographystyle{apsrev}
\bibliography{D:/TEX/Lib/mybib}

\end{document}